\documentclass[10pt,epsfig,aps,floatfix,nofootinbib,superscriptaddress]{revtex4}
\usepackage{graphics}
\usepackage{graphicx}
\usepackage[english]{babel}
\usepackage{fullpage}
\usepackage{amssymb}
\usepackage{amsmath}
\usepackage{verbatim}
\usepackage{pstricks}
\usepackage{slashed}
\usepackage{array}

\newcommand{\mx}{m_{X}}
\newcommand{\ax}{\alpha_X}

\newcommand{\gev}{{\rm GeV}}

\newcommand{\GeV}{{\rm GeV}}
\newcommand{\MeV}{{\rm MeV}}
\newcommand{\mphi}{m_{\phi}}

\newcommand{\xb}{\bar{X}}

\newcommand{\gl}{g_X}
\newcommand{\gr}{g_f}

\newcommand{\vrel}{v_{\text{rel}}}
\newcommand{\sigv}{\langle \sigma v \rangle}
\newcommand{\nn}{\nonumber}

\newcommand{\mpl}{M_\text{pl}}

\newcommand{\nx}{n_X}
\newcommand{\nxb}{n_{\bar{X}}}

\newcommand{\omegac}{\Omega_{\text{CDM}}}
\newcommand{\heff}{h_{\text{eff}}}
\newcommand{\geff}{g_{\text{eff}}}
\newcommand{\eq}[1]{Eq.~(\ref{#1})}
\newcommand{\fig}[1]{Fig.~(\ref{#1})}

\begin{document}

\setlength{\baselineskip}{0.17in}

\begin{flushright}MCTP-11-36\\
\end{flushright}

\vspace{0.2cm}

\title{On Symmetric and Asymmetric Light Dark Matter}

\author{Tongyan Lin}
\email{tongyan@physics.harvard.edu}
\affiliation{Physics Department, Harvard University, Cambridge, MA 02138, USA}
\author{Hai-Bo Yu}
\email{haiboyu@umich.edu}
\affiliation{Michigan Center for Theoretical Physics, Department of Physics, University of Michigan, Ann Arbor, MI 48109\\}
\author{ Kathryn M. Zurek}
\email{kzurek@umich.edu}
\affiliation{Michigan Center for Theoretical Physics, Department of Physics, University of Michigan, Ann Arbor, MI 48109\\}

\date{\today}

\begin{abstract}
\noindent
\fontsize{10}{12}\selectfont

We examine cosmological, astrophysical and collider constraints on
thermal dark matter (DM) with mass $m_{X}$ in the range $\sim1~{\rm
MeV}-10~{\rm GeV}$. Cosmic microwave background (CMB) observations,
which severely constrain light symmetric DM, can be evaded if the DM
relic density is sufficiently asymmetric.  CMB constraints require the
present anti-DM to DM ratio to be less than $\sim2\times10^{-6}$
($10^{-1}$) for DM mass $\mx=1$ MeV (10 GeV) with ionizing efficiency
factor $f\sim1$.  We determine the minimum annihilation cross section
for achieving these asymmetries subject to the relic density
constraint; these cross sections are larger than the usual thermal
annihilation cross section.  On account of collider constraints, such
annihilation cross sections can only be obtained by invoking light
mediators. These light mediators can give rise to significant DM
self-interactions, and we derive a lower bound on the mediator mass
from elliptical DM halo shape constraints. We find that halo shapes
require a mediator with mass $\mphi \gtrsim 4\times10^{-2} \mbox{
MeV}$ ($40 \mbox{ MeV}$) for $\mx=$1 MeV (10 GeV).  We map all of
these constraints to the parameter space of DM-electron and DM-nucleon
scattering cross sections for direct detection. For DM-electron
scattering, a significant fraction of the parameter space is already
ruled out by beam-dump and supernova cooling constraints.

\end{abstract}

\maketitle
\newpage

\section{Introduction}

Studies of dark matter (DM) have historically focused on particles
with weak scale mass $\sim 100\ \gev$
\cite{Bertone:2004pz,Jungman:1995df,Feng:2010gw}. The reason is not
only the focus of the high energy physics community on weak scale
phenomena, but also because the annihilation cross section for a
Weakly Interacting Massive Particle (WIMP) naturally gives
rise to the observed cold DM relic abundance. This is the so-called
``WIMP miracle.''

More recently there has been a broader interest in light DM, with mass
$m_X \lesssim 10\ \gev$.  Part of the reason for this interest is
phenomenological.  Direct detection results from
DAMA~\cite{Bernabei:2010mq},
CoGeNT~\cite{Aalseth:2010vx,Aalseth:2011wp}, and
CRESST~\cite{Angloher:2011uu} claim event excesses that can be
interpreted as nuclear scattering of DM with mass $\sim 10\ \GeV$
(although the mutual consistency of these results is
disputed). Meanwhile dark matter with masses of $\MeV$ has been
studied as a possible explanation of the INTEGRAL 511 keV line
\cite{Boehm:2003ha,Borodatchenkova:2005ct,Hooper:2007tu,Huh:2007zw,Pospelov:2007mp,Hooper:2008im,deNiverville:2011it}.

There is also a theoretical motivation for light DM, as DM with mass
$m_X \lesssim 10\ \gev$ appears in certain classes of models
naturally.  In supersymmetric hidden sector models, for example, gauge
interactions generate light DM masses and give rise to the correct
annihilation cross section
\cite{Hooper:2008im,Feng:2008ya,Feng:2008mu}.  The asymmetric DM (ADM)
scenario, where the DM particle $X$ carries a chemical potential,
analogous to the baryons, provides another approach to light DM~(see
{\it e.g.} \cite{Nussinov:1985xr,Kaplan:1991ah,Barr:1990ca,Kaplan:2009ag}
and references therein). In these
scenarios, both DM ($X$) and anti-DM ($\bar X$) particles may populate
the thermal bath in the early Universe; however, the present number
density is determined not only by the annihilation cross section, but
also by the DM number asymmetry $\eta_X$. Depending on the value for
$\eta_X$, the DM mass can be as low as $\sim {\rm keV}$ in ADM models
\cite{Falkowski:2011xh}, though the natural scale for ADM is set by 
$(\omegac/\Omega_b) m_p\approx 5 \mbox{ GeV}$.

The purpose of this paper is to explore model-independent constraints
and predictions for the asymmetric and symmetric limits of light DM
with mass $\sim1$ MeV$-$10 GeV.\footnote{For DM much lighter than
$\sim$1 MeV, DM can only annihilate to neutrinos, new light states
that remain relativistic through matter-radiation equality, or hidden
sector forces that decay invisibly. In this case, the CMB and collider
bounds discussed here do not apply.}  Although both phenomenological
and theoretical considerations have motivated the study of light DM
candidates, there are still a number of important constraints that
should be taken into account in realistic model building.  In general,
light thermal DM faces two challenges: one is to evade bounds on
energy injection around redshifts $z\sim100-1000$ coming from
observations of the CMB; the other is to achieve the required
annihilation cross section without conflicting with collider physics
constraints.

CMB data from WMAP7 strongly limits DM annihilation during the epoch
of recombination, and excludes symmetric thermal light DM with mass
below $\sim1-10$ GeV if the annihilation is through $s$-wave processes
\cite{Galli:2011rz,Hutsi:2011vx,Finkbeiner:2011dx}. The CMB bounds may
be evaded in the symmetric case if DM dominantly annihilates to
neutrinos or if its annihilation is $p$-wave suppressed. When the DM
relic density is asymmetric, DM annihilation during recombination can
be highly suppressed if the symmetric component is sufficiently
depleted, providing a natural way to resolve the tension from CMB
constraints for light DM scenarios. Unlike the case of symmetric DM,
the CMB places a {\em lower} bound on the annihilation cross section
for ADM from the requirement of sufficient depletion of the symmetric
component.  We calculate the minimum annihilation cross section
required in order to evade the CMB bound and achieve the correct relic
density simultaneously.

However, it is difficult to achieve the needed annihilation rate to
Standard Model (SM) particles through a weak-scale mediator. Null
results from mono-jet plus missing energy searches at the
Tevatron~\cite{Goodman:2010yf,Goodman:2010ku,Bai:2010hh} and the
LHC~\cite{Rajaraman:2011wf,Fox:2011pm} strongly constrain such a
mediator if DM couples to quarks and gluons. Meanwhile, the
mono-photon plus missing energy search at LEP sets limits on the
coupling between DM and charged leptons~\cite{Fox:2011fx} via such a
heavy state. These collider constraints are so strong that the
annihilation through an off-shell heavy mediator is generally
insufficient for ADM to achieve the correct relic density and evade
the CMB constraint, if the DM mass is below a few GeV.  One way to
evade the collider constraints is to invoke a light mediator with mass
much less than $\sim100$ GeV. In this case, DM can annihilate to SM
states efficiently via the light state without conflicting with
collider bounds. Furthermore, if the mediator is lighter than the DM,
a new annihilation channel opens and DM can annihilate dominantly to
the mediator directly.  In this limit, the mediator particle may
couple to the SM sector rather weakly.

The presence of the light mediator has various implications for DM
dynamics in galaxies and for cosmology.  The light mediator may give
rise to significant DM self-interactions (i.e., DM-DM scattering);
this is true in both the symmetric and asymmetric limits, since the
light state mediates DM-DM interactions as well as anti-DM and DM
interactions. These interactions leave footprints in the DM halo
dynamics.  There are limits on the DM self-interaction cross section
coming from observations of elliptical DM halos and elliptical galaxy
clusters. We combine these with the relic
density constraint to place a lower bound on the mediator mass
$\sim4\times10^{-2}~{\rm MeV}-40~{\rm MeV}$ for DM masses in the range
$\sim1~{\rm MeV}-10~{\rm GeV}$.  We assume this massive mediator
decays to SM relativistic degrees of freedom in the early universe to
avoid the overclosure problem, and derive conditions for
thermalization of the DM and SM sectors.

These astrophysical and cosmological constraints can be applied to the
parameter space of scattering rates in direct detection
experiments. We consider DM-nucleon scattering for DM masses of $1-10\
\GeV$ and DM-electron scattering for DM masses $1\ \MeV - 1\ \GeV$.
In the case of electron scattering, we combine the astrophysical and
cosmological constraints with bounds from beam dump experiments and
supernova cooling, which exclude a large region of the allowed
parameter space. In addition, the predictions are very different
dependent on whether the mediator is heavier or lighter than the DM.

The rest of paper is organized as follows. In Section II, we present
the relic density calculation for DM in the presence of a chemical
potential. In Section III, we study the CMB constraint on ADM models
and derive the annihilation cross section required to evade the CMB
bound. In Section IV, we examine current collider physics constraints
on the DM annihilation cross section. In Section V, we study the
elliptical halo shape constraint on the mediator mass. In Section VI,
we map out the parameter space for DM direct detection. We conclude in
Section VII.


\section{Relic Density for Symmetric and Asymmetric Dark Matter \label{sec:cosmo}}

Our starting point is to establish that the correct relic density of
$\omegac h^2 = 0.1109 \pm 0.0056$ \cite{Larson:2010gs} can be
obtained, where we assume that the annihilation cross section $\sigv$
and the asymmetry $\eta_X$ are floating parameters. 

In the usual thermal WIMP scenario, the correct relic density is
determined by DM annihilation until freeze-out. For Dirac DM in the
symmetric limit, the cold DM relic density is $ \omegac h^2 \sim
0.11\left(6\times10^{-26} \text{cm}^3/\text{s}\right)/{\sigv} .  $
DM may also carry a chemical potential which leads to an asymmetry
between the number density of DM and anti-DM. In this case, when the
DM sector is thermalized, the present relic density is determined both
by the annihilation cross section and the primordial DM asymmetry
$\eta_X \equiv (n_X-n_{\bar{X}})/s$, where $n_X$, $n_{\bar{X}}$ are
the DM and anti-DM number densities and $s$ is the entropy density.
In the asymmetric limit, neglecting any washout or dilution
effects, the correct relic density is obtained for a primordial
asymmetry given by
\begin{equation}
  \eta_X \approx \frac{\omegac}{m_X} \frac{\rho_c}{s_0},
  \label{eq:adm}
\end{equation}
where $s_0\approx 2969.5~{\rm cm^{-3}}$ and
$\rho_c\approx1.0540h^2\times10^4~{\rm eV/cm^3}$ are the entropy
density and critical density today. In the asymmetric limit, the
annihilation cross section is sufficiently large that the
thermally-populated symmetric component is a sub-dominant component of
the energy density today.

Depending on the strength of indirect constraints on DM annihilation,
light DM scenarios must interpolate between the symmetric and
asymmetric limits.  We thus require precise calculations of the
present anti-DM to DM ratio $r_\infty = \Omega_{\bar X}/\Omega_X$,
which controls the size of indirect signals from DM annihilation. Note
that $r_\infty$ is related to the absolute relic densities by
\begin{align}
  \Omega_X = \frac{1}{1-r_\infty} \frac{\eta_X m_X s_0}{\rho_c}, \ \
  \Omega_{\bar X} = \frac{r_\infty}{1-r_\infty} \frac{\eta_X m_X s_0}{\rho_c}, \ \
  \label{eq:omegar_relation}
\end{align}
and the total CDM relic density is $\omegac = \Omega_X +
\Omega_{\bar{X}}$.

To compute $r_\infty$ we solve the Boltzmann equations for $n_X,
n_{\bar X}$ freezeout in the presence of a nonzero chemical
potential~\cite{Scherrer:1985zt}.  In this work, we focus on the case
where DM is in thermal equilibrium with the photon thermal bath
through freezeout.  In general, this assumption may not hold if there
is a weakly coupled light mediator coupling DM to the SM. We leave the
more general case for future work \cite{Lin:20XX}, noting that the
effects on the relic density are up to ${\cal O}(10),$ depending on
$m_X$.

The coupled Boltzmann equations for the species $n_+ = n_X$ and $n_- =
\nxb$ are
\begin{equation}
  \frac{dn_\pm}{dt}=-3 H n_\pm- \sigv \left[n_+ n_- - n_+^{eq}n_-^{eq}\right]
  \label{eq:boltzmann}
\end{equation}
where $\sigv$ is the thermally-averaged annihilation cross section
over the $X$ and $\bar{X}$ phase space
distributions~\cite{Gondolo:1990dk}. The Hubble expansion rate is
$H\approx1.66\sqrt{g_{\rm eff}}T^2/\mpl$ where $\mpl\approx 1.22
\times 10^{19} \GeV$ is the Planck mass and $g_{\rm eff}$ is the
effective number of degrees of freedom for the energy density.  If
there is a primordial asymmetry in $X$ number, then there is a nonzero
chemical potential $\mu$ which appears in the equilibrium
distributions as $n_\pm^{eq} = e^{\pm \mu/T} n^{eq}$. Here $n_{eq}$ is
the usual equilibrium distribution with $\mu=0$, and thus
$n_+^{eq}n_-^{eq} = (n^{eq})^2$.

We then take the standard definitions $x=\mx/T$ and $Y_\pm = n_\pm
/s$, where $s=(2\pi^2/45)\heff(T)T^3 $ is the entropy density and
$\heff(T)$ is the effective number of degrees of freedom for the
entropy density. We write the annihilation cross section as
$\sigv=\sigma_0x^{-n}$, with $n=0$ and $n=1$ for $s$-wave and $p$-wave
annihilation processes respectively. Then simplifying
\eq{eq:boltzmann} gives
\begin{align}
  \frac{dY_\pm}{dx} &= - \frac{\lambda}{x^{n+2}} \sqrt{g_*} \left( Y_+ Y_-  - (Y^{eq})^2 \right),
  \label{eq:boltzmannY}
\end{align}
where $\lambda\equiv 0.264 \mpl\mx\sigma_0$ and
$Y^{eq}\simeq0.145(g/\heff)x^{3/2}e^{-x}\equiv a x^{3/2}e^{-x}$.  The
effective number of degrees of freedom is $\sqrt{g_*} =
\frac{\heff}{\sqrt{\geff}} \left( 1 + \frac{T}{3 \heff}
\frac{d\heff(T)}{dT} \right)$ \cite{Gondolo:1990dk}.

After being generated at some high temperature, the DM asymmetry is a
conserved quantity, so we have the constraint
\begin{equation}
	\eta_X = Y_+ -Y_-
	\label{eq:etaXconstraint}
\end{equation}
which is constant at any given epoch.\footnote{We assume there is no
Majorana mass term for DM, and thus $X-\bar X$
oscillation~\cite{Falkowski:2011xh,Cohen:2009fz,Cirelli:2011ac,Buckley:2011ye}
does not occur. We also assume there is no entropy production in this
case and there are no DM-number violating interactions at these
temperatures.} In order to impose this condition on our numerical
solutions, we define the departure from equilibrium $\Delta$ by $Y_\pm
= Y_\pm^{eq} + \Delta$, and instead solve the (single) equation for
$\Delta$.

It is helpful to present approximate analytic solutions in the limit
of constant
$\sqrt{g_*}$~\cite{Scherrer:1985zt,Graesser:2011wi,Iminniyaz:2011yp}.
\eq{eq:boltzmannY} can be solved analytically at late times when
$(Y^{eq})^2$ becomes negligible. In this limit, using
\eq{eq:etaXconstraint}, we can integrate \eq{eq:boltzmannY} separately
for $\bar{X}$ and $X$ to obtain
\begin{align}
  Y_\pm (\infty)&\simeq \frac{\pm \eta_X}{1-\left[1 \mp\eta_X/{Y_\pm}(x_f)\right]e^{\mp \eta_X\lambda \sqrt{g_*} x^{-n-1}_f/(n+1)}}.
  \label{eq:boltzmann_latetime}
\end{align}
These solutions also apply for the symmetric case in the limit of
$\eta_X\to 0$.  We take the freezeout temperature $x_f=\mx/T_f$ as
derived in~\cite{Graesser:2011wi}:
\begin{equation}
  x_f\simeq\ln\left[(n+1)\sqrt{g_*}a\lambda\right]+\frac{1}{2}\ln\frac{\ln^2\left[(n+1)\sqrt{g_*}a\lambda\right]}{{\ln^{2n+4}\left[(n+1)\sqrt{g_*}a\lambda\right]-(\sqrt{g_*})^2\left[(n+1)\lambda\eta_X/2\right]^2}}.
\end{equation}
Using $Y_\pm(\infty)$ given in \eq{eq:boltzmann_latetime}, we can
obtain the present ratio of the $\bar X$ to $X$ number densities:
\begin{equation}
  r_\infty \equiv \frac{Y_-}{Y_+}(\infty) \simeq \frac{Y_-(x_f)}{Y_+(x_f)} \exp \left( \frac{ -\eta_X\lambda \sqrt{g_*}}{ x_f^{n+1} (n+1) } \right).
  \label{eq:rinfty}
\end{equation}
While we can obtain a precise analytic result for $r(x_f) =
Y_-(x_f)/Y_+(x_f)$, it turns out that the consequence of neglecting
the $(Y^{eq})^2$ in the late-time solution can almost exactly be
accounted for by simply setting $r(x_f)=1$. This gives numerically
accurate answers over a wide range of $\eta_X$ and $\sigv$ as
discussed in \cite{Graesser:2011wi}. Note that the solution here only
converges when $\eta_X\lambda$ is small enough $\sqrt{g_*}\eta_X
\lambda < 2 x_f^{n+2}$.


\section{CMB Constraints \label{sec:cmb}}

For both symmetric and asymmetric thermal DM, the DM particles must
have a sufficiently large annihilation cross section in order to
achieve the correct relic density. This annihilation may have many
indirect astrophysical signatures; among these, the most robust
prediction (or constraint) is the effect of DM annihilation on the
cosmic microwave background (CMB) \cite{Padmanabhan:2005es}, since the
effect only depends on the average DM energy density.  We first
summarize recent studies of CMB constraints on DM annihilation, and
then discuss scenarios which naturally evade these constraints for
light DM, focusing on the asymmetric DM scenario.

Energy deposition from DM annihilation distorts the surface of last
scattering, which affects the CMB anisotropies and is thus constrained
by WMAP7 data. CMB constraints become increasingly severe for smaller
DM masses: the energy released in DM annihilations scales as $\sim m_X
(n_X)^2 \sim \rho_{\rm CDM}^2/m_X$, where $\rho_{\rm CDM}$ is the
average energy density in DM. This implies the effect of DM
annihilation on the CMB scales as $\sim \sigv/m_X$. Though the precise
bound depends on the mass and annihilation channels, WMAP7 limits the
amount of annihilation during recombination to below the thermal relic
annihilation cross section if $m_X \lesssim 1-10\ \gev$
\cite{Slatyer:2009yq,Galli:2009zc,Galli:2011rz,Hutsi:2011vx}. Furthermore,
Planck data can improve these constraints by up to a factor of 10.

For self-annihilating DM particles such as
Majorana fermions or real scalars, the energy deposition rate per
volume at redshift $z$ is
\begin{equation}
  \frac{dE}{dt dV}(z) =\rho^2_c \omegac^2 (1+z)^6 f(z) 
        \frac{\sigv_{\rm CMB}}{\mx},
  \label{eq:energyinj_majorana}
\end{equation}
where $\rho_c$ is the critical density at the present time,
$\sigv_{\rm CMB}$ is the thermally-averaged annihilation cross section
at the epoch of recombination, and $f(z)$ parametrizes the amount of
energy absorbed by the photon-baryon fluid at redshift $z$, relative
to the total energy released by DM annihilation at that redshift.

The quantity $f(z)$ gives the efficiency of energy deposition at
redshift $z$ and thus depends on the spectrum of photons, neutrinos
and $e^\pm$ resulting from DM annihilation. In general, the dependence
of $f(z)$ on $z$ is mild \cite{Slatyer:2009yq}, and an excellent
approximation is to take $f(z) \equiv f e_{\rm WIMP}(z)$ where $f$ is
a constant and $e_{\rm WIMP}(z)$ is a universal function for WIMP DM
\cite{Finkbeiner:2011dx}. In addition, to leading order $f \simeq
(1-f_\nu)$ \cite{Hutsi:2011vx}, where $f_\nu$ is the fraction of
energy going to neutrinos per annihilation. For DM annihilation
channels to charged lepton or pion final states, $f \approx 0.2-1$;
here annihilation only to $e^\pm$ can give $f \sim 1$.

There is also some mild $m_X$ dependence in $f(z)$ (or $f$), since the
spectrum of DM annihilation products depends on
$m_X$. Ref.~\cite{Slatyer:2009yq} computed detailed efficiency curves
$f(z)$ for $m_X > 1-10\ \GeV$, depending on the channel. However, the
observed trend is that efficiency does not depend strongly on mass in
the range 1-1000 GeV, and furthermore {\it increases} for lower
mass.\footnote{Above $m_X,m_\phi > 1\ \MeV$, most of the annihilation
products rapidly cascade down to lower energies and the efficiency $f$
is only mildly sensitive to the initial energy spectrum of
annihilation products (normalizing for the total energy). However,
photons in the range $\sim 0.1-1\ \GeV$ deposit their energy
relatively inefficiently.  For annihilation of sub-GeV scale DM,
typically a smaller fraction of the total energy goes into photons of
these energies, which increases the total efficiency slightly. We
thank Tracy Slatyer for this point.} We will extrapolate results to
$m_X < 1\ \GeV$; we expect this is a conservative approach.


\begin{figure}
     \begin{center}
     \includegraphics[width=.75\textwidth]{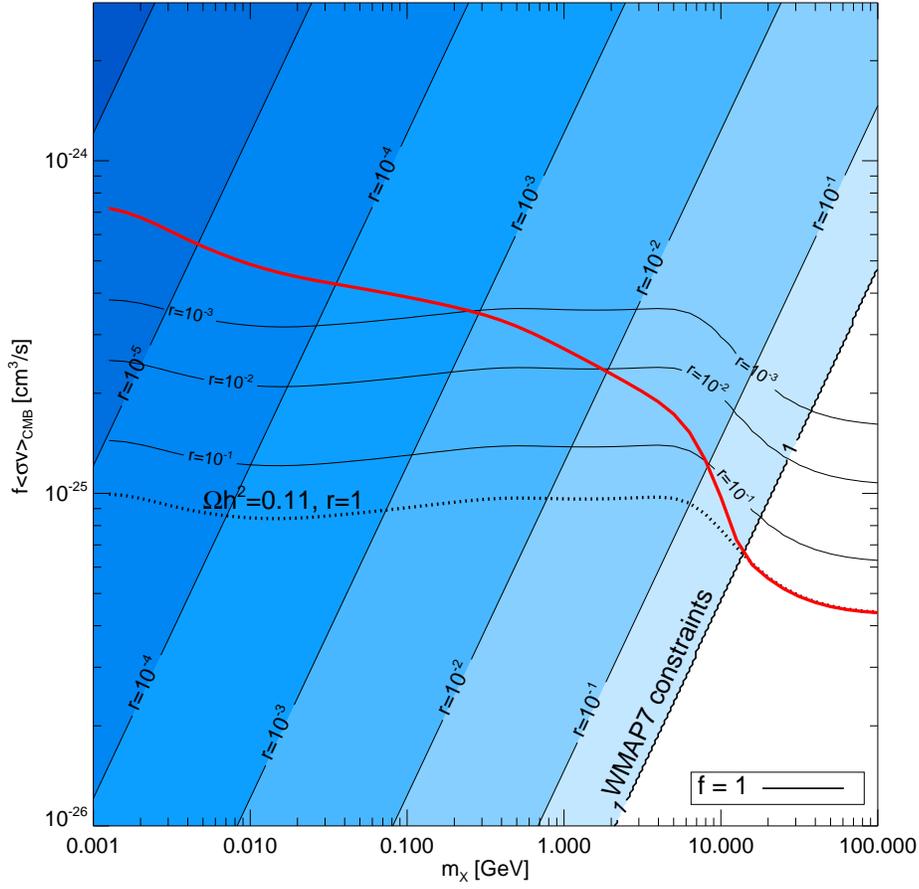}
     \caption{\label{fig:minsigma_largewidth} WMAP7 95$\%$
       C.L. constraints on the DM annihilation cross section and mass
       for asymmetric dark matter and $s$-wave annihilation. We show
       constraints for various values of $r=r_\infty = \Omega_{\bar
       X}/\Omega_{X}$, the anti-DM to DM ratio at the present time.
       The shaded region (blue) is excluded by the WMAP7 data, with
       different shades corresponding to different $r_\infty$.  Along
       the horizontal contours of constant $r$ are the values of
       $\sigv$ where the correct relic density can be obtained for an
       efficiency factor $f=1$. The turnover around $m_X \sim 10\
       \GeV$ comes from the drop in SM degrees of freedom when the
       universe has temperature $\sim1\ \GeV$.  The solid red line is
       the intersection of the WMAP7 and relic density contours: it
       indicates the minimum $\sigv$ needed to obtain the observed relic
       density and satisfy CMB constraints for $s$-wave annihilation.
       }
    \end{center}
\end{figure}


\begin{figure}
     \begin{center}
     \includegraphics[width=.75\textwidth]{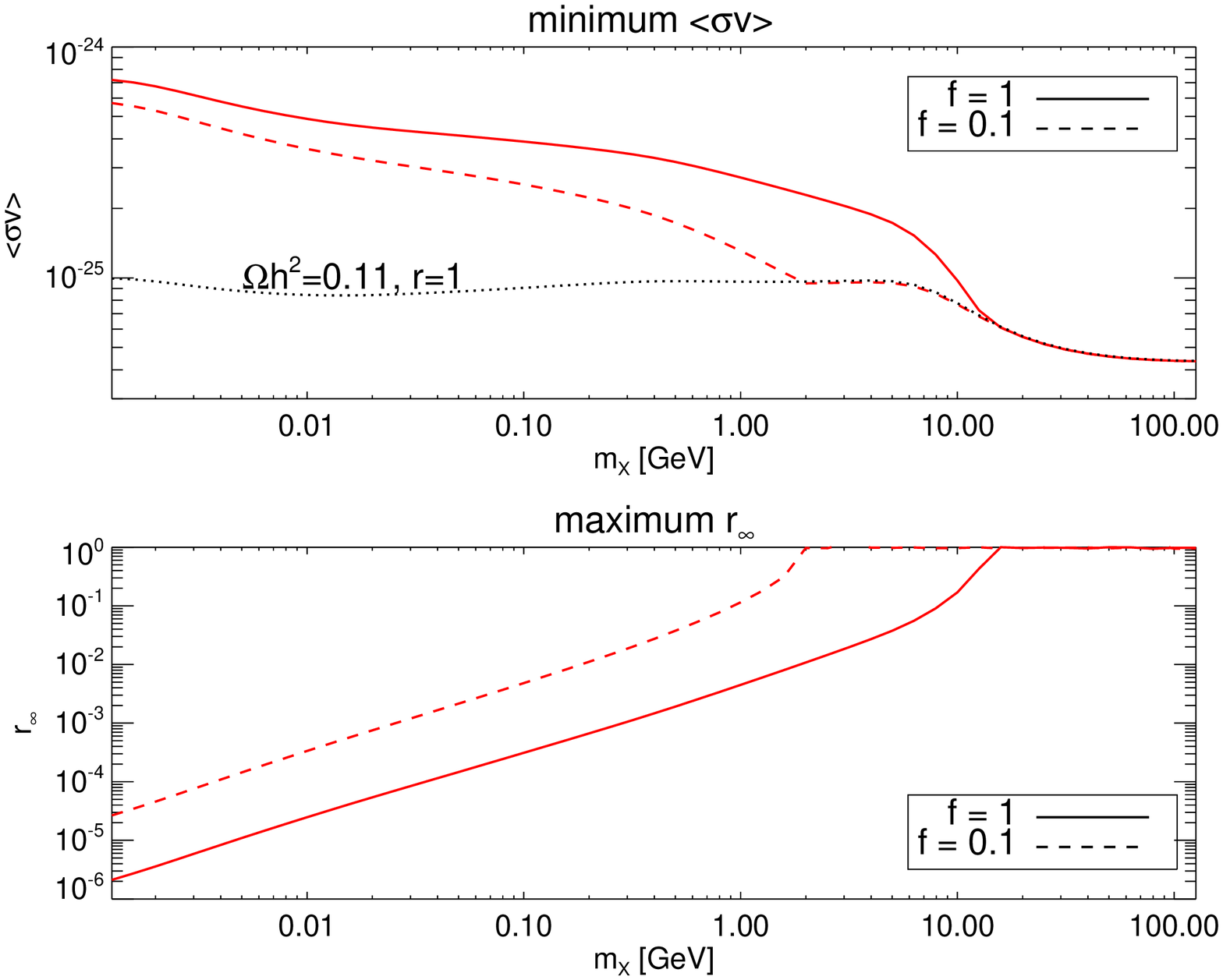}
     \caption{\label{fig:minsigma_largewidth2} (\textit{Top}) Minimum
       $\sigv$ for efficient annihilation of the symmetric component
       in an ADM scenario, such that CMB bounds can be evaded, for two
       different values of the efficiency $f$. The black dotted line
       gives the thermal relic $\sigv$ for the symmetric case.
       (\textit{Bottom}) The corresponding maximum allowed $r_\infty$,
       the anti-DM to DM ratio at the present time.}
       \label{fig:cmb2}
    \end{center}
\end{figure}


The WMAP7 limit on DM energy injection at the $95\%$ C.L. can be
written as \cite{Galli:2011rz}
\begin{equation}
  f\frac{\sigv_{\rm CMB}}{\mx}<\frac{2.42\times10^{-27}~{\rm cm^3/s}}{\GeV}.
  \label{eq:cmb_wmap7}
\end{equation}
This bound\footnote{Note: the results of \cite{Hutsi:2011vx} are
slightly weaker by a factor of 1.2-2.} as given assumes DM particles
are self-annihilating, {\em i.e.} Majorana fermions or real scalars. For DM
candidates that are Dirac fermions or complex scalars, as in ADM
scenarios, the energy injection rate is
\begin{equation}
  \frac{dE}{dt dV}(z) = 2\rho^2_c \omegac^2 
     \frac{r_\infty}{(1+r_\infty)^2}(1+z)^6f(z)\frac{\sigv_{\rm CMB}}{\mx},
  \label{eq:energyinj_dirac}
\end{equation}
where we have used $\rho_X+\rho_{\bar{X}}=\rho_\textrm{CDM}$ and
$r_\infty=\rho_{\bar{X}}/\rho_{X}$. Note there is factor of $2$ in the
energy injection rate relative to the self-annihilating case,
accounting for the number of possible annihilations. Comparing
Eq.~(\ref{eq:energyinj_majorana}) and Eq.~(\ref{eq:energyinj_dirac}),
we can translate the bound given in Eq.~(\ref{eq:cmb_wmap7}) to the
Dirac fermion or complex scalar case:
\begin{equation}
  \frac{2r_\infty}{(1+r_\infty)^2}f\frac{\sigv_{\rm CMB}}{\mx}<
  \frac{2.42\times10^{-27}~{\rm cm^3/s}}{\GeV}.
  \label{CMBadm}
\end{equation}
We show this constraint for various $r_\infty$ values in
\fig{fig:minsigma_largewidth}; the dotted black
line gives the thermal relic annihilation cross section in the
symmetric case, where we have solved for the relic density
numerically and taken $f=1$.

ADM can evade CMB bounds while still allowing $s$-wave
annihilation.\footnote{In the symmetric limit, one can evade the CMB
bounds if DM annihilates via $p$-wave suppressed interactions. Then
\mbox{$\sigv_{\rm CMB}~\simeq~(v_{\rm CMB}/v_f)^2~\sigv_f$} and since $v_{\rm
CMB}\sim10^{-8}$ while $v_f\sim0.3$, the annihilation cross section at
recombination is highly suppressed and WMAP constraints are
substantially weakened. An increased branching ratio to neutrinos
(smaller $f$) can also alleviate the tension with CMB data for light
DM.}  The CMB bounds do not completely disappear in the ADM scenario,
however, because there is a small symmetric component of DM remaining,
$r_\infty$, the size of which depends on $\sigv$.  Because of the
exponential dependence of $r_\infty$ on $\sigv$, as shown in
Eq.~(\ref{eq:rinfty}), the CMB constraints lead to a lower bound on
$\sigv$.  This is shown in \fig{fig:minsigma_largewidth}, where we map
out the constraints in the $\sigv_\textrm{CMB}$ and $m_X$ parameter
space, computing the relic density numerically and applying the
constraint in \eq{eq:cmb_wmap7}.  The solid line (red) gives the
resulting lower bound on $f \sigv_\textrm{CMB}$. This lower bound on
$f \sigv_\textrm{CMB}$ translates to an upper bound on the residual
symmetric component, $r_\infty$, as shown in
\fig{fig:minsigma_largewidth2}.  We give analytic approximations to
these numerical solutions next.

When $r_\infty\ll 1$, we can ignore the $\bar X$ contribution to the
total relic density, and the DM asymmetry parameter $\eta_X$ is set by
$\eta_X\approx \Omega_{\rm CDM}\rho_c/(\mx s_0)$.  For a given
$\eta_X$, the required annihilation cross section at freezeout to
achieve a particular residual symmetric component, $r_\infty$, can be
obtained by rewriting Eq.~(\ref{eq:rinfty}) as
\begin{align}
  \sigv_f&\simeq \frac{s_0x_f}{0.264\omegac \rho_c\sqrt{g_{*,f}} \mpl }
            \ln\left(\frac{1}{r_\infty}\right) \nn \\
  &\simeq c_f \times 5\times10^{-26}~{\rm cm^3/s} \times 
            \ln\left(\frac{1}{r_\infty}\right),
  \label{relicasy}
\end{align}
where $c_f \equiv
\left(\frac{x_f}{20}\right)\left(\frac{4}{\sqrt{g_{*,f}}}\right)$ is
an ${\cal O}(1)$ factor. We show the numerical result as the
horizontal contours of constant $r_\infty$ in
\fig{fig:minsigma_largewidth}; for $m_X < 1\ \GeV$ we obtain a
good approximation to the numerical solution by taking $c_f = 1$.  On
the other hand, the CMB bound on the annihilation cross section when
$r_\infty\ll 1$ is
\begin{equation}
  \sigv_{\rm CMB}<\frac{2.42\times10^{-27}~{\rm cm^3/s}}{2f}\left(\frac{\mx}{1\ \GeV}\right)\left(\frac{1}{r_\infty}\right).
  \label{cmbasy}
\end{equation}
For $s$-wave annihilation, we take $\sigv_f \simeq \sigv_{\rm CMB}$.
Since $\sigv_f$ increases with $\log (1/r_\infty)$, but the CMB bound
on $\sigv_{\rm CMB}$ increases with $1/r_\infty$, we can evade the CMB
constraints by decreasing $r_\infty$. For a given DM mass, thermal ADM
is consistent with the CMB constraints if $r_\infty$ satisfies the
following condition,
\begin{align}
  r_\infty\ln\left(\frac{1}{r_\infty}\right)<\frac{2.42\times10^{-2}}{f \times c_f} \left(\frac{\mx}{1~{\rm
GeV}}\right).
  \label{eq:rbound}
\end{align}
The numerical result for this bound is shown in
Fig.~(\ref{fig:minsigma_largewidth2}); a good analytic approximation is
given by $r_\infty < r_0 / \ln \left( 1/r_0 \right)$, with $r_0 \simeq 2
\times 10^{-2} (m_X/\GeV) /f$.  Taking $f\sim1$, we can see that
$r_\infty$ has to be smaller than $5\times 10^{-3}$ and
$2\times10^{-6}$ for $\mx\sim 1\ \GeV$ and $1\ \MeV$, respectively.

Likewise, we can combine \eq{relicasy} and \eq{cmbasy}
to place a lower bound on $\sigv_f$:
\begin{align}
  \frac{\sigv_f}{c_f \times 5\times10^{-26}~{\rm cm^3/s}} \gtrsim 
  \begin{cases}
    \ln \left( 40 c_f f \times \frac{1\ \GeV}{m_X} \right) + \ln \ln  \left( 40 c_f f \times \frac{1\ \GeV}{m_X} \right) & , \ \ m_X \lesssim f \times 10 \GeV. \\
    2 & , \ \ m_X \gtrsim f \times 10 \GeV.
  \end{cases}
  \label{eq:sigmabound}
\end{align}
Note if $m_X$ is larger than $ f\times 10\ \GeV$, the CMB constraints
do not apply and the annihilation cross section is set by the relic
density requirement.  The analytic approximation in \eq{eq:sigmabound}
agrees well with the numerical results, which are shown in
\fig{fig:minsigma_largewidth2}.

With these constraints on the minimum annihilation cross section, we
now turn to discussing what classes of models can generate the needed
annihilation cross section consistent with collider constraints.


\section{Light Mediators \label{sec:motivation}}

Thus far, we have treated the annihilation cross section $\sigv$ as a
free parameter. To proceed we must specify the physics that generates
this cross section. First, DM may annihilate directly to SM particles
through heavy mediators with mass greater than the weak scale. This
coupling to the SM implies light DM can be produced in abundance in
colliders. We review constraints from missing (transverse) energy
searches at collider experiments and from direct detection
experiments, which conflict with the $\sigv$ required to obtain the
observed relic density. In this case, thermal light DM is ruled
out in both the symmetric and asymmetric scenarios. Second, DM can
annihilate via new light states which have a mass below the typical
momentum transfer scale in the colliders. In this case, the collider
constraint can be evaded. If the new state is lighter than DM, it can be very weakly coupled to the SM.


\subsection{Collider and Direct Detection Constraints on Light DM with Heavy Mediators}

In the heavy mediator case, a convenient way to parametrize the DM-SM
coupling is via higher dimensional operators, which is valid if the
mediator mass is heavier than the relevant energy scale.  Here we give
two typical examples,
\begin{equation}
  {\cal O}_1:\frac{\bar{X}\gamma_\mu X\bar{f}\gamma^\mu f}{\Lambda^2_1}~{\rm and}~
  {\cal O}_2:\frac{\bar{X}X\bar{f}f}{\Lambda^2_2},
\end{equation}
where $X$ is DM, $f$ is a SM fermion, and $\Lambda_{1,2}$ are cut-off
scales for ${\cal O}_{1,2}$. The cut-off scale, in terms of the
parameters in the UV-complete models, is
$\Lambda=\mphi/\sqrt{\gl\gr}$, where $\mphi$ is the mediator mass, and
$\gl$ and $\gr$ are coupling constants of DM-mediator and SM-mediator
interactions respectively.

In the limit of $\mx\gg m_f$, the DM annihilation cross sections at
freezeout are given by
\begin{equation}
  \sigv_1\simeq\frac{N^c_f}{\pi}\frac{\mx^2}{\Lambda^4_1}~{\rm and}~\sigv_2\simeq\frac{N^c_f}{8\pi}\frac{\mx^2}{\Lambda^4_2}\frac{1}{x_f}, 
\end{equation}
for ${\cal O}_1$ and ${\cal O}_2$ respectively. $N^c_f$ is the color
multiplicity factor of fermion $f$, and $x_f=\mx/T \approx 20$, with
$T$ the temperature.  Note that the annihilation cross section through
${\cal O}_2$ is $p$-wave suppressed.  Now we can estimate the limit on
the cut-off scales $\Lambda_1$ and $\Lambda_2$ by requiring the
correct relic density
\begin{align}
\label{lambda1}\Lambda_1&\lesssim 370~\GeV \left(\frac{N^c_f}{3}\right)^{\frac{1}{4}}\left(\frac{\mx}{10~\GeV}\right)^{\frac{1}{2}}\left(\frac{6\times10^{-26}~{\rm cm^3/s}}{\sigv}\right)^{\frac{1}{4}},\\
\label{lambda2}\Lambda_2&\lesssim 100~\GeV \left(\frac{N^c_f}{3}\right)^{\frac{1}{4}}\left(\frac{\mx}{10~\GeV}\right)^{\frac{1}{2}}\left(\frac{6\times10^{-26}~{\rm cm^3/s}}{\sigv}\right)^{\frac{1}{4}}\left(\frac{20}{x_f}\right)^{\frac{1}{4}},
\end{align}
where the limit is relevant for both the asymmetric and symmetric
cases.  Since the annihilation cross section is $p$-wave suppressed
for ${\cal O}_2$, we need a smaller cut-off scale to obtain the
correct relic abundance. Now we review various constraints on the
cut-off scales $\Lambda_{1,2}$.

\begin{itemize}
\item{\bf Direct Detection Constraints} 

  If DM couples to quarks, the operators ${\cal O}_{1,2}$ can lead to
  direct detection signals with the DM-nucleon scattering cross
  section: $\sigma_{n_{1,2}}\sim\mu^2_n/\Lambda^4_{1,2}$, and $\mu_n$
  is the DM-nucleon reduced mass.  For a DM mass $\sim 10~{\rm GeV}$,
  taking the value of $\Lambda_{1,2}$ given in Eqs.~(\ref{lambda1})
  and (\ref{lambda2}), we expect the DM-nucleon scattering cross
  section to be $\sigma_{n_1}\sim10^{-38}~{\rm cm^2}$ and
  $\sigma_{n_2}\sim10^{-36}~{\rm cm^2}$. However, the current upper
  bound on $\sigma_n$ from direct detection experiments for DM with
  mass $m_X \gtrsim 10 \mbox{ GeV}$ is $\sigma_n\lesssim10^{-42}~{\rm
  cm^2}$ \cite{Aprile:2011hi}, which is much smaller than the
  predicted values from requiring the correct thermal relic density.
  For DM with mass below a few $\gev$, the recoil energies are too
  small and direct detection bounds are currently very weak or
  nonexistent.

\item{\bf Tevatron and LHC Constraints}

  The DM-quark interactions given in ${\cal O}_{1,2}$ can lead to
  signals of mono-jet plus missing transverse energy at hadron
  colliders, while the Tevatron data for this signal matches the SM
  prediction well. We require that ${\cal O}_{1,2}$ do not give rise
  to sizable contributions to this signal. The lower bounds on
  $\Lambda_{1,2}$ are $\sim 400~{\rm GeV}$ and $\sim 400~{\rm GeV}$
  \cite{Goodman:2010yf,Goodman:2010ku,Bai:2010hh} respectively, for
  DM masses $m_X \lesssim 10~{\rm GeV}$ that we are interested in.
  Recent LHC results give a stronger limit on
  $\Lambda_1\gtrsim700~{\rm GeV}$~\cite{Fox:2011pm}.  Therefore the
  Tevatron and LHC searches have excluded both thermal symmetric DM
  and ADM in the whole range of light DM if the DM particles
  annihilate to light quarks through ${\cal O}_1$ and ${\cal O}_2$.
  
\item{\bf LEP Constraints}

  If DM particles couples to the electron through ${\cal O}_{1,2}$,
  the mono-photon search at LEP sets a limit on the cut-off scale:
  $\Lambda_1\gtrsim480~{\rm GeV}$ and $\Lambda_2\gtrsim440~{\rm GeV}$
  for DM mass $m_X \lesssim 10~{\rm GeV}$ \cite{Fox:2011fx}.  Note the
  limit also applies to the case where DM couples to three generations
  of charged leptons universally.  One may avoid the limit by coupling
  DM only to $\mu$ or $\tau$.  However this approach usually involves
  model building complications and severe flavor constraints.

\end{itemize}

Thus we conclude that for ${\cal O}_{1,2}$, DM does not have the
correct relic abundance for symmetric DM and ADM due to the
combination of direct detection and collider constraints.  The direct
detection constraints can be relaxed by suppressing the direct
detection scattering cross section; this can happen for example if the
scattering off nuclei is velocity suppressed, notably through an axial interaction.  However, the collider bounds are still
severe for higher dimensional operators involving interactions with
light quarks or
electrons~\cite{Goodman:2010yf,Goodman:2010ku,Bai:2010hh,Fox:2011fx,Rajaraman:2011wf,Fox:2011pm}.


\subsection{Light Dark Matter with Light Mediators \label{sec:lightmed}}

One simple way to evade the collider constraints for light DM is to
invoke light mediators with masses much smaller than the typical
transverse momentum of the colliders $p_T\sim{\cal O}(100~{\rm GeV})$
(or the center of mass energy $\sim 200$ GeV for LEP). In this
limit, the effective theory approach breaks down and the collider
bounds become much
weaker~\cite{Bai:2010hh,Fox:2011fx,Fox:2011pm,Graesser:2011vj}.  In
general, if the mediator mass is much less than the $p_T$ probed at
colliders, there exists a large parameter space for light DM scenarios
to achieve the correct relic density.  We consider a hidden sector
with Dirac DM coupled to a light mediator which could be a spin-1 or
spin-0 particle; for ease of notation we always refer to it as
$\phi$. We write the Lagrangians as
\begin{align}
  {\cal L}_V &= g_X\bar{X}\gamma^\mu X\phi_\mu+g_f\bar{f}\gamma^\mu f\phi_\mu+ m_X \bar X X+ m^2_\phi \phi^\mu\phi_\mu, \\
  {\cal L}_S &= g_X\bar{X} X\phi+g_f\bar{f}f\phi + m_X \bar X X +m^2_\phi\phi^2,
  \label{eq:lagrangian}
\end{align}
where $\mphi$ is the mediator mass. We consider two cases for the
mediator mass:\footnote{In this paper, we do not consider the
intermediate case $\mphi\sim2\mx$, where there is a resonance in the
s-channel annihilation of $\bar X X$.} a mediator with $m_\phi > 2
m_X$ and lighter mediator with $m_\phi < m_X$.

In the case of $p_T\gg\mphi>2\mx$, the DM particles can annihilate to
SM particles through the s-channel process.  There is a collider bound
on $g_f$ because an on-shell mediator which decays to $X\bar{X}$ can
be produced, potentially contributing to the mono-jet plus missing
transverse energy signal.  Tevatron data has been employed to place an
upper bound on $g_f<0.015/\sqrt{Br(\phi\rightarrow X\bar{X})}$ for
$\mphi<20~{\rm GeV}$~\cite{Graesser:2011vj}, where $Br(\phi\rightarrow
X\bar{X})$ is the branching ratio of $\phi$ decay to the DM pair. In
this case the annihilation cross section is given by $\sigv_V\simeq
4\alpha_Xg^2_f\mx^2N^c_f/\mphi^4$ and $\sigv_S\simeq
\alpha_Xg^2_f\mx^2N^c_f/2\mphi^4x_f$, where $\alpha_X\equiv g^2_X/4\pi$. To
see how the collider constraint affects the annihilation cross section
in this case, we take the conservative limit $g_f\lesssim0.015$,
setting $Br(\phi\rightarrow X\bar{X})\sim1$. From the relic density
constraint, we then obtain an upper bound on the mediator mass,
\begin{equation}
  m_\phi \lesssim 13~{\rm GeV}\left(\frac{\alpha_X}{10^{-1}}\right)^{1/4}\left(\frac{10^{-25}~{\rm cm^3/s}}{\left<\sigma v\right>}\right)^{1/4}\left(\frac{\mx}{1~{\rm GeV}}\right)^{1/2}.
\end{equation}
This bound\footnote{Note that in this case there are also strong
bounds on $m_\phi$ from neutrino experiments
\cite{deNiverville:2011it}; however, we have checked that it is still
possible to obtain the correct relic density and that the direct
detection predictions are unaffected.} is consistent with our
assumption that $m_\phi \gg m_X$.

If $\mphi<\mx$, DM can annihilate to the mediator directly
and the annihilation cross section is determined primarily by the hidden
sector coupling $g_X$:
\begin{equation}
  \sigv_V=\frac{\pi\alpha^2_X}{\mx^2}\sqrt{1-\left(\frac{\mphi}{\mx}\right)^2}, \ \ \ \sigv_S=\frac{9}{2}\frac{\pi\alpha^2_X}{\mx^2}\frac{T}{\mx}\sqrt{1-\left(\frac{\mphi}{\mx}\right)^2}
  \label{eq:sigv_lightmediators}
\end{equation}
for the vector and scalar mediators respectively. Meanwhile $g_f$
determines how the DM sector couples to the SM sector. As for the
collider physics, the production of $X \bar{X}$ occurs through an
off-shell mediator; since this is a three-body process, the bound is
rather weak. Tevatron data requires $g_f \lesssim 0.2$ if the mediator
couples to quarks universally~\cite{Graesser:2011vj}.

Although $g_f$ does not appear to play an important role in the relic
density, this coupling controls the width (lifetime) of $\phi$ and is
relevant for cosmology.  The width $\Gamma_\phi$ of the mediator is
\begin{equation}
   \left( \Gamma_\phi \right)_V =  \frac{4N^c_f}{3}\frac{m_\phi}{16\pi} 
     g_f^2 \sqrt{1 - \left( \frac{2 m_f}{m_\phi} \right)^2} , \ \ \
   \left( \Gamma_\phi \right)_S =  2N^c_f\frac{m_\phi}{16\pi} g_f^2 \sqrt{1 - \left( \frac{2 m_f}{m_\phi} \right)^2}, 
\end{equation}
where the lifetime $\tau_\phi = \Gamma_\phi^{-1}$.  In
Section~\ref{sec:cosmo}, we assumed the DM particles to be in thermal
equilibrium with the SM thermal bath in the early universe, and in
this case the standard freezeout picture and cosmology apply. Now, we
check the condition for thermalization of the two sectors.  If the
mediator decay rate is larger than the Hubble expansion rate at
temperatures $T > m_\phi$, then the inverse decay processes can keep
$\phi$ in chemical equilibrium with the SM thermal
bath~\cite{Feng:2010zp}. At these temperatures, the decay rate is
given by $\Gamma_\phi \sim {g^2_f\mphi^2}/{(16\pi T)}, $
where the factor of $\mphi/T$ accounts
for the effect of time dilation.  In order for the mediator to stay in
thermal equilibrium with the SM thermal bath through DM freezeout, we
require $\Gamma_\phi \gtrsim H$ at temperatures $ T \sim \mx$. This
gives a constraint on $g_f$:
\begin{align}
  \gr \sim \sqrt{\frac{16\pi\Gamma_\phi}{m_\phi}} \gg 8 \times 10^{-8} \left(\frac{\sqrt{\geff}}{9}\right)^{1/2} \left( \frac{m_X}{\GeV} \right)^{3/2} \left(\frac{100\ \MeV}{\mphi}\right).
  \label{eq:gammaphi2}
\end{align}

If $g_f$ is less than the bound given in Eq.~(\ref{eq:gammaphi2}), the
DM sector can have a different temperature from the SM sector and the
standard freezeout calculation can be modified in a number of ways. We
have checked that these effects lead to change in the minimum
annihilation cross section by less than a factor ${\cal O}(10)$,
compared to the results we derived, in
Sections~\ref{sec:cosmo}-\ref{sec:cmb}.  Furthermore, the massive
mediator is a late-decaying particle and in the case where the
mediator decays to the SM states, can modify standard nucleosynthesis
(BBN).  There are stringent constraints on the hadronic decay of
long-lived particles from the $^4{\rm He}$ fraction, which requires
that the lifetime of the mediator be less than $10^{-2}~{\rm s}$
\cite{Kawasaki:2004qu,Kawasaki:2004yh,Jedamzik:2006xz}.  This leads to
a lower bound of $g_q \gtrsim 1.6 \times 10^{-11} \sqrt{1\
\GeV/m_\phi}$ for a vector mediator, where we take $N^c_f=3$.  For
leptonic decay modes, we take the lifetime of the mediator $\tau_\phi
\lesssim 1~s$, and obtain a slightly weaker bound, $g_e \gtrsim 5
\times 10^{-11} \sqrt{10~{\rm MeV}/m_\phi}$, for a vector mediator
with $N^c_f=1$.

Finally, we comment on the calculation of the relic density and
application of the CMB constraints in the light mediator case. When
$m_\phi < m_X$, $\xb X$ can annihilate to $\phi\phi$, but $\phi$
decays to standard model particles rapidly compared to the relevant
time scales at recombination so that the CMB constraints are unchanged.
The only difference between a heavy mediator and light mediator with
large width is whether there is a contribution to the effective
degrees of freedom, $g_*$, from the light mediator. A slightly higher
$g_*$ in the light mediator case gives rise to smaller $r_\infty$,
which in turn weakens the lower bound on $\sigv$ from CMB constraints.

In addition, we have neglected the Sommerfeld enhancement effect. As
we will discuss in the following section, the mediator mass is bounded from
below by DM halo shapes; this limits the size of any Sommerfeld
enhancement. In addition, since $\sigv \approx \pi \alpha_X^2/m_X^2$,
for light DM the coupling $\alpha_X$ can be much smaller and still
satisfy the relic density constraint. For the DM masses considered
here, we have checked that the Sommerfeld enhancement effect is
negligible for $s$-wave and $p$-wave annihilation processes at both
freezeout and during recombination, if we take $\alpha_X$ and $m_\phi$
close to their minimum allowed values.


\section{Halo Shape Constraints on the Mediator Mass \label{sec:haloshape}}

The presence of the light mediator allows for significant DM
self-interactions, which can have non-trivial effects on DM halo
dynamics. A number of astrophysical observations constrain DM
self-interactions, for example observations of the Bullet
Cluster~\cite{Markevitch:2003at}, elliptical galaxy
clusters~\cite{MiraldaEscude:2000qt} and elliptical DM
halos~\cite{Feng:2009mn,Feng:2009hw}.  Among these, the upper bound on
DM self-interaction from the ellipticity of DM halos is the
strongest~\cite{Feng:2009mn}. DM self-interactions can erase the
velocity anisotropy and lead to spherical DM halos, so the observed
ellipticity of DM halos constrains the DM self-scattering rate.
Because the strength of self-interaction increases as the mediator
mass decreases, we can use the elliptical halo shape constraint to
place a lower limit on the mediator mass. Note that in the case of
$\mphi=0$, the ellipticity of the DM halos then places a strong upper
limit on the hidden sector coupling $g_X$~\cite{McDermott:2010pa}; it
is only possible to obtain the correct relic density if $m_X \gtrsim 10^3\
\GeV$~\cite{Feng:2009mn,Ackerman:2008gi}\footnote{This
limit can be relaxed if the hidden sector is much colder than the
visible sector when DM freezes out. In this case, DM can achieve the
correct relic density with a smaller annihilation cross
section~\cite{Feng:2011uf}.} .

The effect of DM self-interactions on DM halo shapes can be
parametrized by the average rate for DM particles to change velocities
by ${\cal O}(1)$~\cite{Feng:2009hw}:
\begin{equation}
  \Gamma_k=\int d^3v_1d^3v_2f(v_1)f(v_2)(\nx\vrel\sigma_T)(\vrel^2/v^2_0),
  \label{selfrate}
\end{equation}
where $n_X$ is the DM density in the DM halo, $\vrel = |\vec v_1 -
\vec v_2|$, and $f(v)$ is the DM velocity distribution in the DM halo,
for which we take $f(v)=e^{-v^2/v^2_0}/(v_0\sqrt{\pi})^3$. $\sigma_T$
is the scattering cross section weighted by the momentum transfer:
$\sigma_T=\int d\Omega_*(d\sigma/d\Omega_*)(1-\cos\theta_*)$.

The form of $\sigma_T$ depends on the particle physics nature of DM
self-interactions and the relevant momentum scales. If the mediator is
lighter than the typical momentum transfer in collisions, DM particles
interact through long-range forces and $\sigma_T$ depends on
velocity. In the opposite limit where the mediator is heavy compared
to momentum transfer, DM self-interactions are contact interactions
and $\sigma_T$ is independent of $\vrel$.  In this case, we can take
the $\sigma_T$ out of the velocity integrals in Eq.~(\ref{selfrate})
and the calculation is straightforward.  We first will derive the
upper bound on the DM self-interaction cross section assuming a
contact interaction, and then show that this limit applies in deriving
the minimum mediator mass.

\begin{figure}
    \begin{center}
    \includegraphics[width=.75\textwidth]{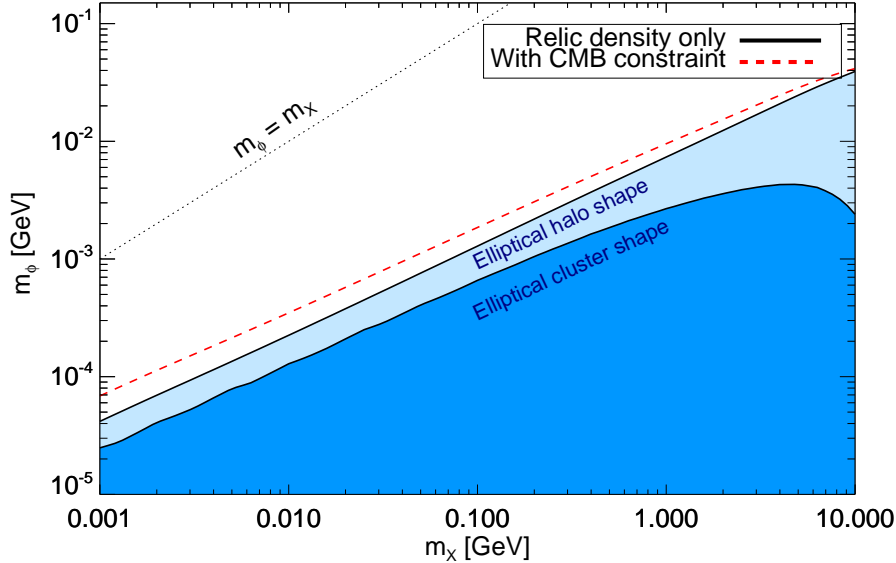}
    \caption{\label{fig:haloshape} Lower limit on the mediator mass
     from combining relic density and DM self-interaction
     constraints. We show the case of a vector mediator; the result
     for a scalar mediator is similar and is given in \eq{massbound}.
     We consider DM self-interaction constraints from elliptical halo
     shapes and elliptical cluster shapes. Bullet cluster constraints
     do not give a lower bound on $m_\phi$. The dashed red line
     indicates the bound on the mass from elliptical halo shapes if
     CMB bounds are also applied, assuming efficiency $f \approx 1$. }
   \end{center}
\end{figure}

We consider the well-studied elliptical galaxy
NGC720~\cite{Buote:2002wd,Humphrey:2006rv}, taking our bound from the
observed ellipticity at a radius of 5 kpc. The DM density profile is
fit with local density $4~{\rm GeV/cm^3}$ and radial velocity
dispersion $\bar{v}_r^2=v^2_0/2\simeq(240~{\rm km/s})^2$. We require the
average time for DM self-interactions to create ${\cal O}(1)$ changes
on DM velocities to be larger than the galaxy lifetime $t_g\sim
10^{10}~{\rm years}$ i.e. $\Gamma^{-1}_k>t_g$. This gives the upper
bound
\begin{equation}
  \sigma_T \lesssim 4.4 \times 10^{-27}~{\rm cm^2}
       \left(\frac{\mx}{1\ \GeV}\right)
       \left(\frac{10^{10}{~\rm years}}{t_g}\right).
  \label{strongestbound}
\end{equation}
The reader should bear in mind that this is an analytic estimate
and detailed N-body simulations studying a range of elliptical
galaxies are required for a robust bound.

Other astrophysical constraints have been derived for $\sigma/m_X$,
assuming a hard sphere scattering cross section $\sigma$. A similar
bound derived from shapes of elliptical galaxy clusters is
($\sigma/m_X \lesssim 10^{-25.5}~{\rm
cm^2}(\mx/\GeV)$)~\cite{MiraldaEscude:2000qt}. Specifically, this
estimate is obtained from the inner regions of the galaxy cluster
MS2137-23, at a radius of 70 kpc with dark matter density $\sim 1\
\GeV/\textrm{cm}^3$. Cosmological simulations of cluster-sized objects
support this estimate within an order of magnitude
\cite{Yoshida:2000uw}; however, the bound is still based on a single
cluster.  There is also a bound derived from the Bullet Cluster
($\sigma/m_X \lesssim 2\times10^{-24}~{\rm
cm^2}(\mx/\GeV)$)~\cite{Markevitch:2003at}, reproduced in simulations
of the collision by \cite{Randall:2007ph}. Note that this result is
not derived from the shapes of the merging clusters but from requiring
that the subcluster does not lose a significant fraction of its mass
in passing through the larger cluster; however, we have found that the
bound is too weak in this case to give a minimum mediator mass.


For the vector and scalar interactions considered here, the force is
described by a Yukawa potential $V(r)=\pm\ax e^{-\mphi r}/r$.
Depending on the mediator, and whether we are in the asymmetric limit,
the sign may be positive or negative. For the vector case, we have
both $XX$ interactions (+) and $X\bar X$ interactions (-) unless we
are in the asymmetric limit. For the scalar case, the sign is always
negative.  However, in the limit of a contact interaction, the sign of
the potential does not matter. 
The momentum transfer cross section for scattering through $t$ and $u$-channel processes in the Born approximation is
\begin{equation}
  \sigma_T \approx \frac{4\pi\alpha^2_X\mx^2}{\mphi^4},
\end{equation}
which is subject to the bound in Eq.~(\ref{strongestbound}).  We have assumed a contact interaction, $\mx\vrel/\mphi \ll 1$; we will justify later that this is a valid assumption in deriving the bounds below.

On the other hand, the relic density constraint places a lower bound
on the annihilation cross section $\sigv \gtrsim 10^{-25}
\textrm{cm}^3/\textrm{s}$ for light DM and thus on $\alpha_X$:
\begin{align}
  \alpha_X{|_V} & \gtrsim 5\times10^{-5}
         \left(\frac{\sigv}{10^{-25}{\rm cm^3/s}}\right)^{1/2}
         \left(\frac{\mx}{\GeV}\right), \nonumber \\
  \alpha_X{|_S} & \gtrsim 11\times10^{-5}
         \left(\frac{\sigv}{10^{-25}{\rm cm^3/s}}\right)^{1/2}
         \left(\frac{\mx}{\GeV}\right)\left(\frac{x_f}{20}\right)^{1/2},
  \label{alphax}
\end{align}
for vector and scalar coupling respectively. Note that we assume
$m_\phi < m_X$ and take the annihilation cross sections in
\eq{eq:sigv_lightmediators}. 

Since $\alpha_X$ cannot be arbitrarily small, $m_\phi$ cannot be made
arbitrarily small. Combining the bound on $\alpha_X$ with
\eq{strongestbound}, we obtain a lower bound on the mediator mass:
\begin{align}
  \mphi|_V & \gtrsim 7\ \MeV 
       \left(\frac{\sigv}{10^{-25}{\rm cm^3/s}}\right)^{1/4}
       \left(\frac{\mx}{\GeV}\right)^{3/4}, \nonumber \\
  \mphi|_S & \gtrsim 11\ \MeV 
       \left(\frac{\sigv}{10^{-25}{\rm cm^3/s}}\right)^{1/4}
       \left(\frac{x_f}{20}\right)^{1/4}\left(\frac{\mx}{\GeV}\right)^{3/4}
\label{massbound}
\end{align}
for the vector and scalar mediator cases, where we take the elliptical
galaxy with $t_g=10^{10}~{\rm years}$.  Note that because the bound on
$\mphi$ scales as $\sigma_T^{-1/4}$ in the contact interaction limit,
the result is not very sensitive to the precise bound on $\sigma_T$.

In deriving the above bound on $m_\phi$, we have assumed that
$\mphi\gg\mx\vrel$ and that the Born approximation is valid.  Now we
check that the bound given in Eq.~(\ref{massbound}) is consistent with
these assumptions. The condition $\mphi\gg\mx\vrel$ is satisfied for
$1\ \MeV < m_X < 10\ \GeV$, since from \eq{massbound} we have
$m_\phi/m_X \sim 10^{-2} (m_X/\GeV)^{-1/4}$ but $\vrel\sim10^{-3}$.
In this limit the Born approximation is valid if the following
condition is satisfied
\begin{equation}
  \mx\left|\int^\infty_0rV(r)dr\right|=\frac{\mx\ax}{\mphi}\ll1.
\label{born}
\end{equation}
From Eq.~({\ref{alphax}}), we can see $\vrel\gg\alpha_X$ in the DM
mass range we are interested in, and thus this condition is also
satisfied if $\mphi\gg\mx\vrel$.  We emphasize that we ${\it cannot}$
extrapolate the lower mass bound given in Eq.~(\ref{massbound}) to
$m_X \gtrsim 50~{\rm GeV}$ because the Born approximation breaks down.
For these higher masses, in general one has to solve the scattering
problem numerically~\cite{Buckley:2009in}.  In the classical limit
where $ \mx\vrel \gg \mphi$, there is a fitting formula available
in~\cite{Khrapak:2003xx} for the transfer cross section, which has
been used to study self-interactions via a light mediator for DM
masses greater than $\sim100~{\rm
GeV}$~\cite{Feng:2009hw,Ibe:2009mk,Feng:2010zp,Loeb:2010gj}.

In \fig{fig:haloshape} we show the lower limit on $m_\phi$ for the
vector case, including the result derived from the more conservative
bounds from elliptical cluster shapes. We also show the slightly
stronger result if we take the CMB constraint on the cross
section,\footnote{In the scalar case, annihilation is $p$-wave
suppressed and thus CMB constraints don't apply.} given in
\eq{eq:sigmabound}.  There is a turnover for the elliptical cluster
bounds because the contact interaction limit breaks down; here we use
the full cross section, again in the Born approximation, given in
\cite{Feng:2009hw}.  The bounds from the Bullet Cluster, which we derive
following \cite{Feng:2009mn}, do not give rise to a lower bound on
$\mphi$.

\section{Direct Detection \label{sec:directdetection}}

Given the experimental effort needed to detect DM directly, it is
important to map out the parameter space of direct detection cross
sections, subject to the astrophysical and cosmological constraints we
have discussed. Current experiments are not sensitive to DM-nucleon
scattering if the DM mass is below $\sim$1 GeV because of the energy
thresholds.  It has been suggested that DM-electron scattering may
provide an alternative way for the detection of light
DM~\cite{Essig:2011nj}.  We consider DM-nucleon scattering for
$\mx\gtrsim1~{\rm GeV}$ and DM-electron scattering for $1~{\rm
MeV}\lesssim\mx\lesssim1~{\rm GeV}$.

We compute the range of allowed elastic scattering cross sections
within the framework of light DM annihilating via hidden sector
mediators, assuming mediator couplings to electrons or light
quarks. We consider both lighter mediators, $m_\phi < m_X$, and
heavier mediators, where we focus on the case $m_\phi \gg m_X$. When
$m_\phi < m_X$ the mediator can be very weakly coupled to the SM, and
so the scattering cross sections can be much smaller than when $m_\phi
\gg m_X$. However, there is still a lower limit on the cross section
coming from the lower bounds on the couplings of the mediator to the
DM and SM fermions, $\alpha_X$ and $g_f$ respectively.  The lower
bound on $\alpha_X$ is derived from requiring that relic density and
CMB constraints are satisfied. We consider two possible lower bounds
on $g_f$: from requiring the thermalization between the DM and SM
sectors, or from requiring decay of the mediator before BBN. When
$m_\phi \gg m_X$ the lower limit on the cross section arises purely
from the relic density and CMB constraints.

Meanwhile, we obtain upper bounds on the electron scattering
cross section from the combination of halo shape bounds and requiring
that the mediator does not significantly affect the electron anomalous
magnetic moment. Including supernova and beam dump constraints on the
dark force coupling \cite{Bjorken:2009mm} then carves out a nontrivial
part of the parameter space for electron scattering.

\begin{figure}
     \begin{center}
     \includegraphics[width=.47\textwidth]{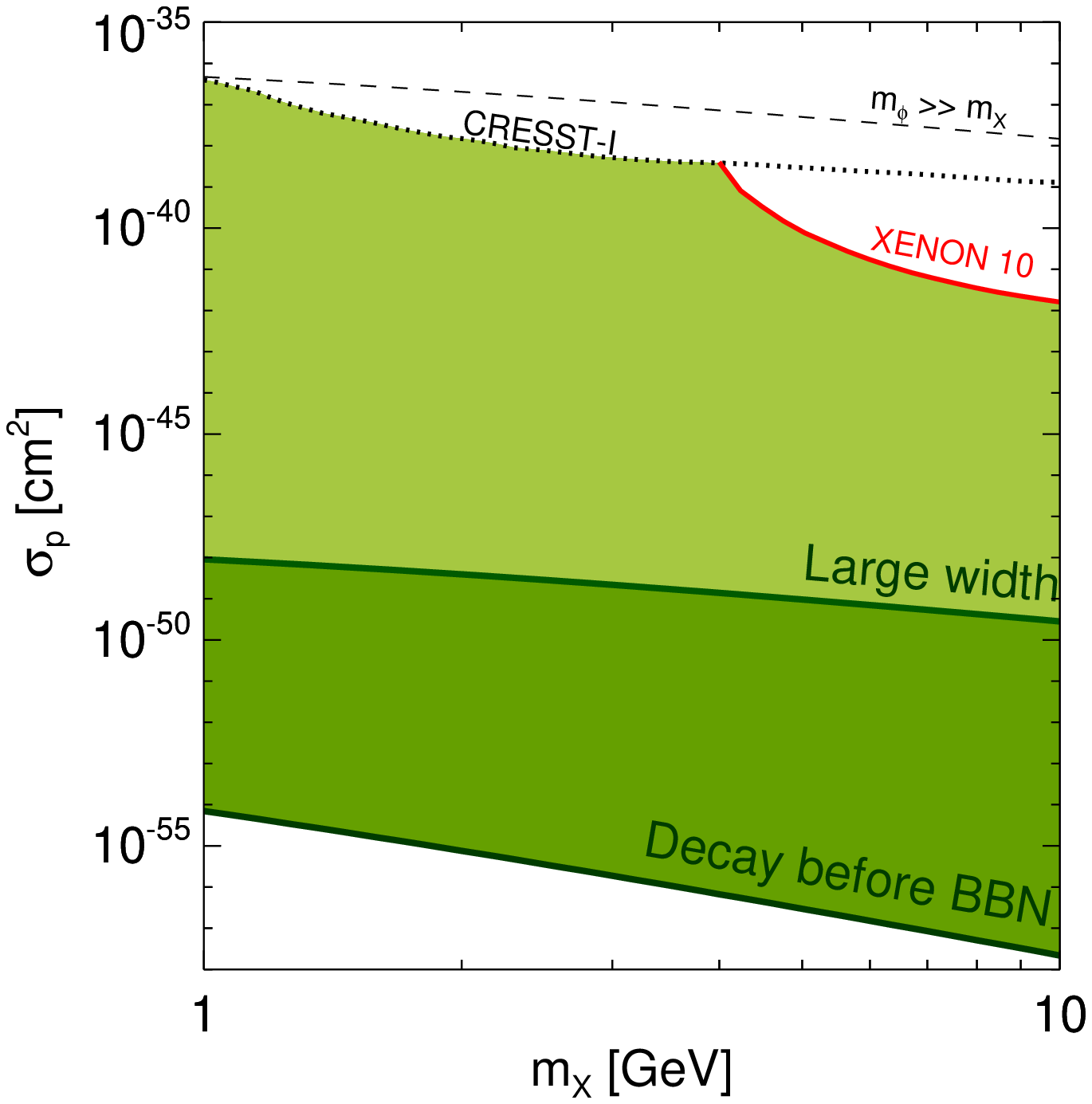}
     \includegraphics[width=.47\textwidth]{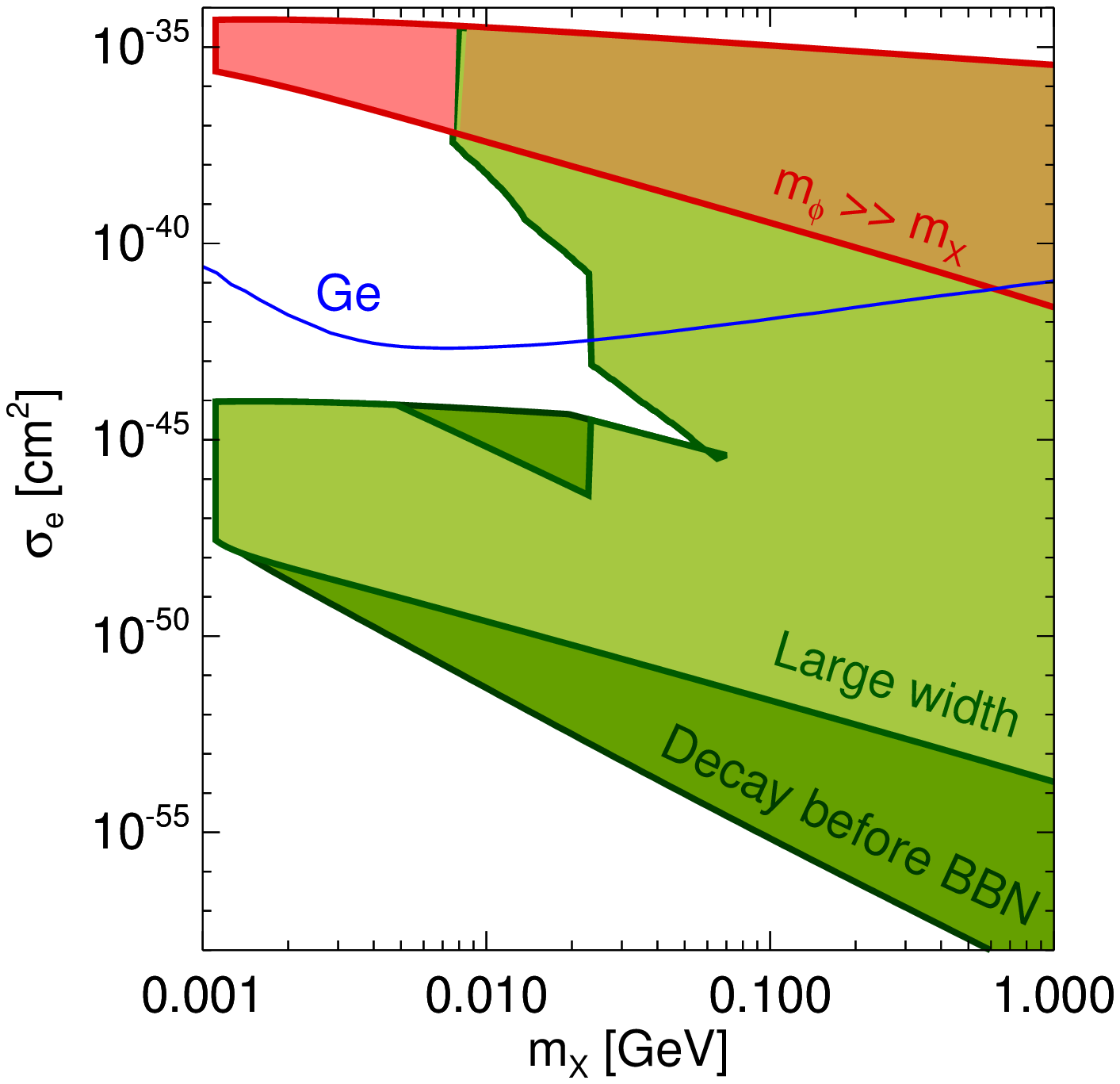}
     \caption{\label{dd_SI} \textit{(Left)} Nucleon scattering through
     a vector mediator. The green shaded region indicates the allowed
     parameter space of direct detection cross sections. The lighter
     green region imposes the bound of thermal coupling between the
     two sectors (``large width'') while the larger shaded region only
     requires mediator decay before BBN. Also shown is the lower bound
     for the heavy mediator ($m_\phi \gg m_X$) case.
     \textit{(Right)} Electron
     scattering through a vector mediator, for
     $m_\phi < m_X$ (green) and  $m_\phi \gg m_X$ (red);
     the intersection of the two regions is shaded brown. We show the
     projected sensitivity of a Ge experiment, taken from
     \cite{Essig:2011nj}. Beam dump, supernova, and halo shape
     constraints apply here and carve out the region of large
     $\sigma_e$ at low $m_X$.
     For more details, see the text. In the lighter green
     region, the condition of thermal equilibrium between the visible and
     hidden sectors is imposed.}
          \end{center}
\end{figure}

Fig.~(\ref{dd_SI}) summarizes our results for the case where the
mediator is a vector. We show the possible DM-nucleon (left panel) and
DM-electron (right panel) scattering cross sections as a function of
DM mass.  The green shaded region is the parameter space for $m_\phi <
m_X$ which is allowed by the constraints from the relic density, BBN,
and DM halo shape constraints; in the electron case we include beam
dump and supernova cooling constraints. The lighter green area is set
by the additional assumption that the mediator has large decay width
and thus that the two sectors are in thermal equilibrium. In the
nucleon scattering case, $m_\phi \gg m_X$ is ruled out by CRESST-I and
XENON10. In the electron scattering case, the red shaded region gives
the allowed cross sections for $m_\phi \gg m_X$. In the following
sections we derive these results and present more details.

\subsection{Nucleon Scattering}
We first consider nucleon scattering in the mass range $1\ \GeV
\lesssim m_X \lesssim 10\ \GeV$, taking universal couplings to the
light quarks given by $g_q$.  The DM-nucleon scattering cross section
is given by
\begin{equation}
  \sigma_n=4\ax g^2_n\frac{\mu^2_n}{\mphi^4},
  \label{Nscattering}
\end{equation}
where $\mu_n$ is the WIMP-nucleon reduced mass, and $g_n = 3 g_q$ is
the $\phi_\mu$-nucleon coupling constant.  The upper bounds here are
set by results from direct detection experiments, in particular
CRESST-I \cite{Angloher:2002in} and XENON10 \cite{Angle:2011th}.  We
have taken a contact interaction; this is a good approximation over
much of the parameter space because the momentum transfer is generally
less than the minimum mediator mass allowed by the ellipticity of DM
halos, as discussed in Section~\ref{sec:haloshape}.  We note that
momentum-dependence can be relevant for scattering off heavier nuclei
such as xenon if we take $m_\phi$ to be close to this minimum value,
and thus can change the upper limit from XENON10
\cite{Fornengo:2011sz,Feldstein:2009tr,Farina:2011pw}. However, the
lower limit is obtained in the limit that $m_\phi \approx m_X$ and
thus momentum dependence will not be important. We therefore consider
the bounds on a contact interaction for simplicity.

To determine the lower limit on this cross section, we bound
$\alpha_X$ and $g_q$ from below in the case that the mediator is
lighter than the DM, $m_\phi < m_X$.  For thermal DM and masses $m_X >
1\ \GeV$, a lower bound on $\alpha_X$ is determined primarily by the
relic density.  As described in Section~\ref{sec:cmb}, CMB constraints
are only important in this mass range if $\phi_\mu$ decays dominantly
to electrons, for which the efficiency factor is $f\sim 1$.  For
$\phi_\mu$ coupling primarily to quarks, $f\approx 0.2$ and CMB bounds
don't apply above $m_X \sim 2\ \GeV$.
Then the minimum annihilation cross section is $\sigv \approx \pi
\alpha_X^2/m_X^2 \approx 10^{-25}$cm$^3$/s, giving a bound of
$\alpha_X \gtrsim 5.2 \times 10^{-5} (m_X/\GeV)$.  Requiring thermal
equilibrium between the hidden and visible sectors, we take the bound
on $g_q$ in \eq{eq:gammaphi2}, with $\sqrt{\geff} \approx 9$.
Combining the limits above results in a lower bound on the nucleon
scattering cross section:
\begin{equation}
  \sigma_n \gtrsim 10^{-48} \textrm{cm}^2 \times \left(\frac{\mx}{\GeV}\right)^4 \left(\frac{\GeV}{\mphi}\right)^6 \left(\frac{\mu_n}{0.5 \GeV}\right)^{2}.
  \label{eq:sigmaN_bound_thermal}
\end{equation} 
Since $m_\phi < m_X$, this quantity is saturated for any $m_X$ if we
set $m_\phi$ to its maximum value of $m_\phi \sim m_X$.  This bound is
indicated by the ``Large width'' line in
Fig.~(\ref{dd_SI}). Coincidentally, the lower limit here is similar to
the best achievable sensitivity for WIMP-nucleon scattering if the
dominant irreducible background is coherent scattering of atmospheric
neutrinos off of nuclei
\cite{Strigari:2009bq,Monroe:2007xp,Freedman:1973yd}. However, these
studies focused on WIMP DM; for light DM, solar
neutrinos become much more important and the best achievable
sensitivity may be several orders of magnitude weaker.

The lower bound on $\sigma_n$ given in
Eq.~(\ref{eq:sigmaN_bound_thermal}) is derived by requiring the two
sectors be in thermal equilibrium.  We may relax this assumption, and
just demand the mediator decay by nucleosynthesis. This gives $g_q
\gtrsim 1.6 \times 10^{-11} \sqrt{1\ \GeV/m_\phi}$, as discussed in
Section~\ref{sec:lightmed}. For such $g_q$ the two sectors are
decoupled through freezeout; then the relic density calculation is
slightly more complicated and depends on the thermal history of the
sectors.  The change in the relic density then modifies the bound on
$\alpha_X$.  We have checked that the full calculation generally only
changes the bound on $\alpha_X$ by an ${\cal O}(1)$
factor~\cite{Lin:20XX}, so here we take the bound on $\alpha_X$
from the large $\phi$ width case for simplicity.  
In this limit, the lower bound on $\sigma_n$ is given by
\begin{equation}
  \sigma_n \gtrsim 5 \times 10^{-54} \textrm{cm}^2 \times \left(\frac{\mx}{\GeV}\right) \left(\frac{\GeV}{\mphi}\right)^5 \left(\frac{\mu_n}{0.5 \GeV}\right)^{2}
  \label{eq:sigmaN_bound}
\end{equation} 
labeled as ``Decay before BBN'' in Fig.~(\ref{dd_SI}).

For reference, we also give the lower bound on the cross section in
the case where $m_\phi \gg m_X$.  Here DM annihilation occurs
directly to SM final states through $\phi_\mu$, with annihilation
cross section $\sigv = 4\alpha_X g_n^2 m_X^2/m_\phi^4$. Since the same
combination of parameters enters in both the annihilation cross
section and the nucleon scattering cross section, we can directly
apply the relic density constraint to obtain
\begin{equation}
  \sigma_n \gtrsim 5 \times 10^{-37} \mbox{ cm}^2  \left(\frac{1~{\rm GeV}}{m_X}\right)^{2}\left(\frac{\mu_n}{0.5~{\rm GeV}}\right)^{2}.
\end{equation}
This is the ``$m_\phi \gg m_X$'' line in \fig{dd_SI}.
However, this scenario is ruled out by the direct detection limits on
the cross section.


\subsection{Electron Scattering}

\begin{figure}
     \begin{center}
     \includegraphics[width=.47\textwidth]{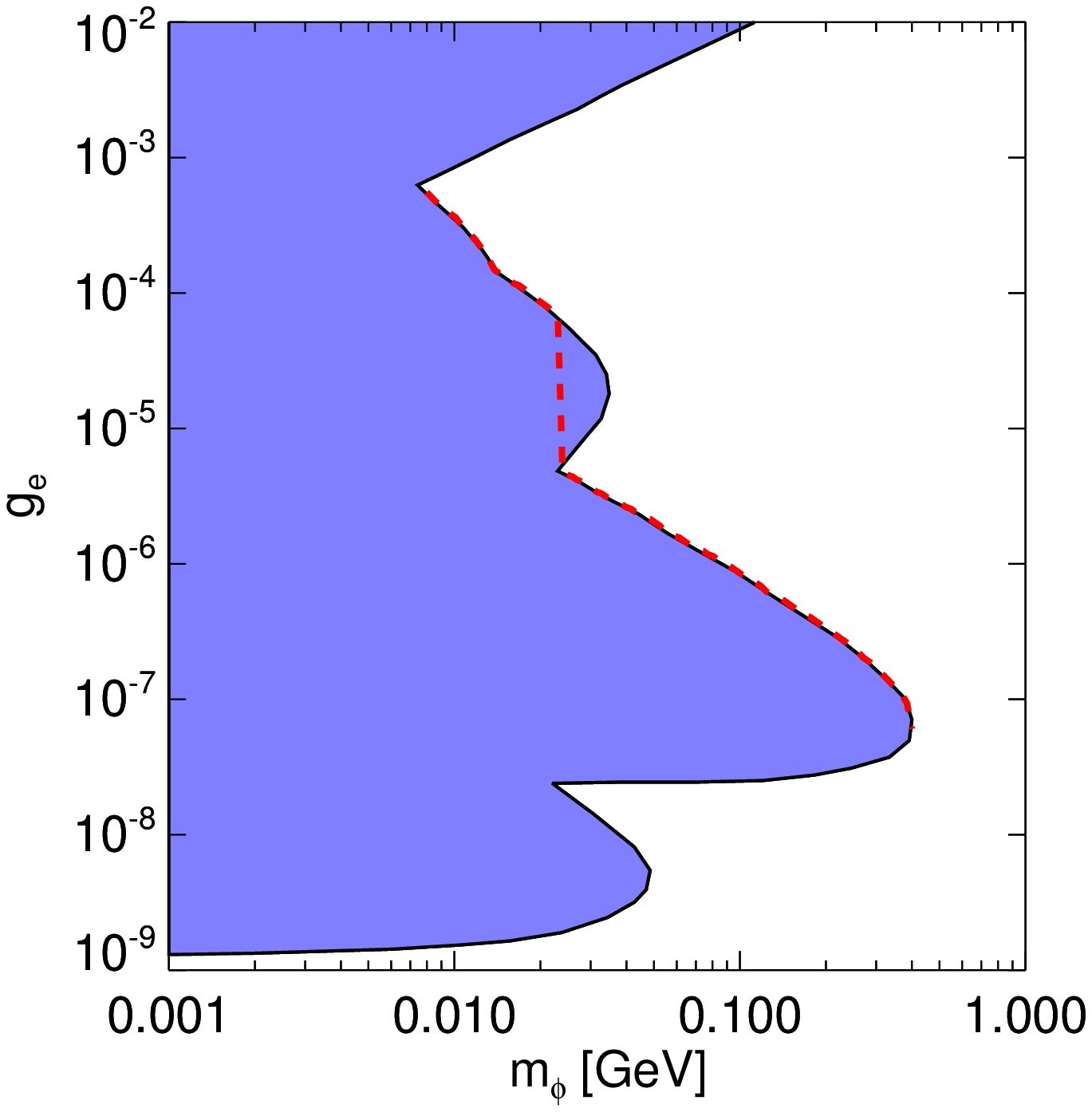}
     \includegraphics[width=.47\textwidth]{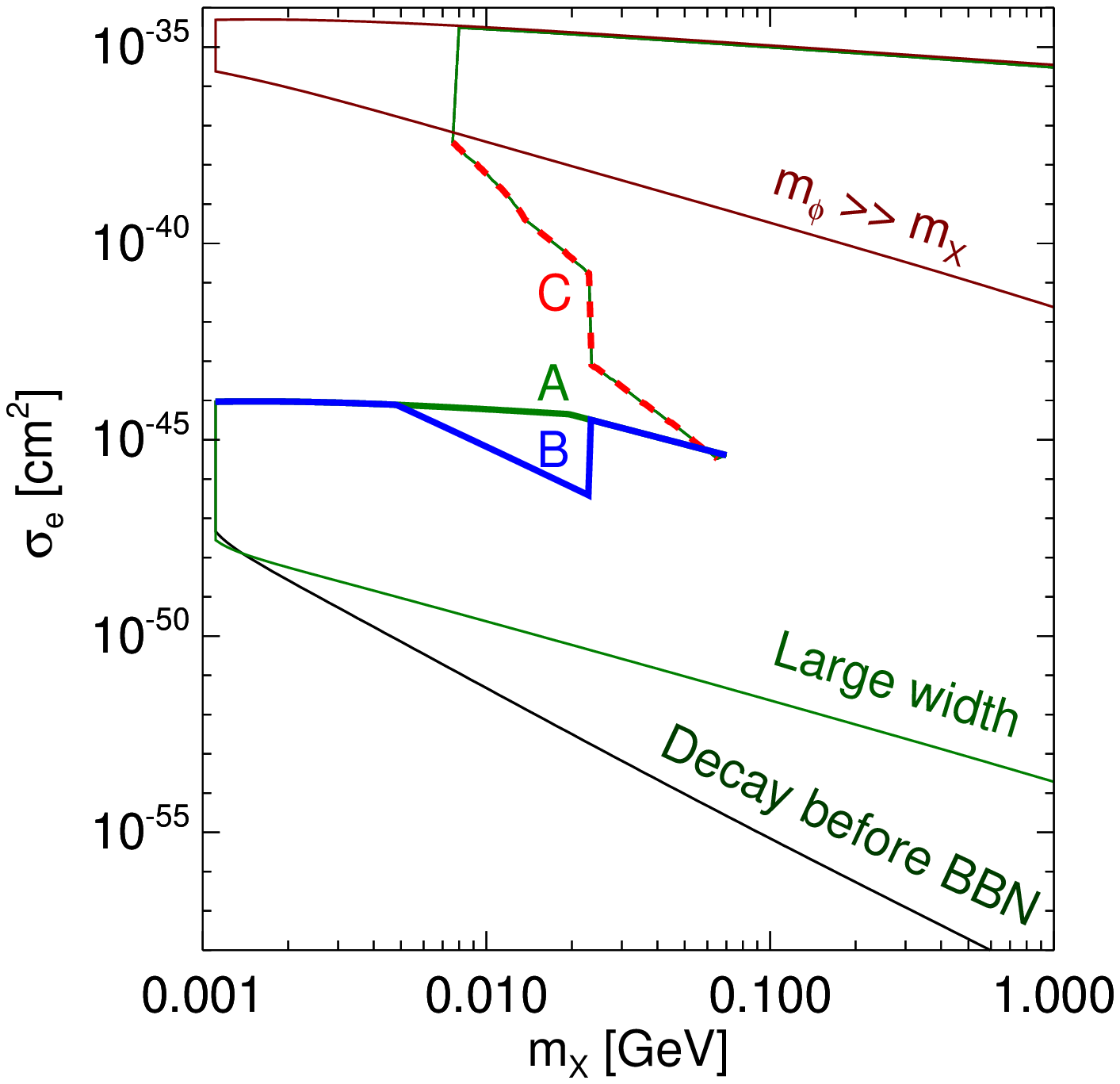}
     \caption{\label{dd_electron_detail} \textit{(Left)} Constraints
       on mediator mass $m_\phi$ and coupling to electrons $g_e$ for
       $m_\phi < m_X$. The shaded region is excluded from electron
       anomalous magnetic moment, beam dump experiments, and supernova
       cooling \cite{Bjorken:2009mm}.  The red dashed line shows the
       $g_e$ value used to derive the corresponding red dashed line
       (``C'') in the right plot.  \textit{(Right)} Constraints on
       electron scattering from Fig.~\ref{dd_SI}. The boundaries A, B,
       and C are discussed in more detail in the text.}
     \end{center}
\end{figure}

We consider scattering off electrons for DM in the mass range $1\ \MeV
< m_X < 1\ \GeV$. The DM-electron scattering cross section is
\begin{equation}
  \sigma_e=4\ax g^2_e\frac{\mu^2_e}{\mphi^4}.
  \label{eq:e_scattering}
\end{equation}

The lower bound on the scattering cross section can be derived in the
same way as in the nucleon case, taking $m_\phi < m_X$.  Here both CMB
and relic density constraints apply, since $m_X < 1\ \GeV$ and the
energy deposition efficiency $f \approx 1$ for decay to electrons. We
take the bound on the annihilation cross section in \eq{eq:sigmabound}
with $c_f \approx 1$, giving a lower limit on $\alpha_X$:
\begin{equation}
  \alpha_X \gtrsim 4 \times10^{-7} \left( \frac{m_X}{10\ \MeV} \right) \sqrt{ \ln \left( \frac{40\ \GeV}{m_X} \right) }.
  \label{eq:electron_minalphaX}
\end{equation}

As in the nucleon case, a lower bound on the DM-electron
scattering cross section can be derived by assuming that the hidden
and visible sectors are in thermal equilibrium.  Analogously to
\eq{eq:sigmaN_bound_thermal}, we find
\begin{equation}
  \sigma_e \gtrsim 3 \times 10^{-51} \textrm{cm}^2 \times 
     \left(\frac{\mx}{10\ \MeV}\right)^4
     \left(\frac{10\ \MeV}{\mphi}\right)^6
     \left(\frac{\mu_e}{0.5\ \MeV}\right)^{2} 
\sqrt{ \ln \left( \frac{40\ \GeV}{m_X} \right) },
  \label{eq:sigmae_bound_thermal}
\end{equation} 
where we take $\sqrt{g_{\rm eff}} \approx 3$.

Again, it is possible that the DM sector thermal bath evolves
independently from the SM sector and in this case we only require the
mediator to decay before BBN. From Section~\ref{sec:lightmed}, we take
the bound $g_e\gtrsim5\times10^{-11}\sqrt{10~{\rm MeV}/\mphi}$.  The
minimum scattering cross section is
\begin{equation}
  \sigma_e \gtrsim 3 \times10^{-53}~{\rm cm^2}
     \left(\frac{\mx}{10\ \MeV}\right)
     \left(\frac{10\ \MeV}{\mphi}\right)^5
     \left(\frac{\mu_e}{0.5\ \MeV}\right)^{2}
  \sqrt{ \ln \left( \frac{40\ \GeV}{m_X} \right) }.
  \label{eq:sigmaE_bound}
\end{equation} 
If the annihilation goes through a heavier mediator $m_\phi \gg m_X$,
we derive the strongest lower bound on the scattering cross section by
applying CMB and relic density constraints:
\begin{equation}
  \sigma_e \gtrsim 4 \times 10^{-39} \mbox{ cm}^2  
          \left(\frac{10\ \MeV}{m_X}\right)^{2}
          \left(\frac{\mu_e}{0.5\ \MeV}\right)^{2}
  \ln \left( \frac{40\ \GeV}{m_X} \right) .
\end{equation}

For electron scattering there are no direct experimental bounds on
$\sigma_e$. However, for $m_\phi < m_X$, there are bounds on
$\sigma_e$ arising from indirect constraints, namely halo shape bounds
and from searches for new light gauge bosons
\cite{Bjorken:2009mm}. The halo shape constraint requires that the
self-scattering cross section satisfy $\sigma_T/m_X < 4.4 \times
10^{-27} \textrm{cm}^2/\GeV$ with $\sigma_T \simeq 4\pi \alpha_X^2
m_X^2/m_\phi^4$. If $m_\phi < m_X$ then constraints on new light gauge
bosons rule out parts of the $(m_\phi,g_e)$ parameter space; we show
beam dump, supernova cooling and electron anomalous magnetic moment
constraints\footnote{In general there are also constraints from
low-energy $e^+e^-$ colliders, fixed target experiments, and neutrino
experiments \cite{deNiverville:2011it}. 
We find these do not significantly affect our results.  
In the case of kinetic mixing, bounds from measurements of the muon
anomalous magnetic moment also apply. We do not include them in this
paper.}  in Fig.~(\ref{dd_electron_detail}) (left panel). Here we make
use of the convention in \cite{Bjorken:2009mm}, where $g_e = \epsilon
e$, with the kinetic mixing parameter $\epsilon \equiv \epsilon_Y \cos
\theta_W$ and $e$ electric charge. The solid line (and shaded region)
indicates the constraint.

As a simple application of the constraints discussed above, we derive
the upper bound on the cross section by rewriting $\sigma_e$:
\begin{align}
  \sigma_e &= \frac{4 \mu_e^2}{\sqrt{4 \pi m_X}} \sqrt{\frac{\sigma_T}{ m_X}} \left(\frac{g_e}{m_\phi} \right)^2 \nn \\
  &\lesssim 3.5 \times 10^{-35} \mbox{ cm}^2 
          \left( \frac{\mu_e}{0.5\ \MeV} \right)^2  
          \sqrt{ \frac{10\ \MeV}{m_X} } .
\end{align}
Here we have applied the halo shape constraint and taken
$(g_e/m_\phi)^2 \lesssim 10^{-1} e^2/\GeV^2$, arising from
measurements of the electron anomalous magnetic moment
\cite{Pospelov:2008zw}.

To explain more complicated constraints on the $(\mx$,$\sigma_e)$
plane from the supernova cooling and beam dump experiments for $m_\phi
< m_X$, we show again the allowed parameter space for electron
scattering cross sections, but highlight boundaries of the constraints
by labeling ``A'', ``B'', and ``C'' in the right panel of
Fig.~(\ref{dd_electron_detail}). We can map excluded regions on the
$(\mphi,g_e)$ plane to these constraints:
\begin{itemize}
  \item \textbf{Constraint ``A'':}

    For $m_\phi < m_X \lesssim 8\ \MeV$, supernova plus beam dump
    constraints require $g_e \lesssim 1.3 \times 10^{-9}$. This places
    a stringent upper bound on the cross section, which we derive by
    taking $m_\phi$ to its minimum value of $m_\phi = 2 m_e \approx 1\
    \MeV$, and then setting $\alpha_X$ to the maximum value allowed by
    halo shape constraints: $\alpha_X < 9.5 \times 10^{-6} \sqrt{10\
    \MeV/m_X}$. This upper bound is then
    \begin{equation}
      \sigma_e \lesssim 6 \times 10^{-45} \mbox{ cm}^2 
          \left( \frac{\mu_e}{0.5\ \MeV} \right)^2  
          \sqrt{ \frac{10\ \MeV}{m_X} }.
    \end{equation}
    Note that the constraint changes somewhat if we also consider
    $m_\phi < 1\ \MeV$. In this case, supernova cooling constraints
    still require $g_e \lesssim 1.3 \times 10^{-9}$ but halo shapes
    allow for a somewhat smaller $m_\phi$. As a result, the upper
    bound is slightly weaker if we allow $m_\phi < 1\ \MeV$: $
    \sigma_e \lesssim 6 \times 10^{-44} \mbox{ cm}^2 \left( \mu_e/0.5\
    \MeV \right)^2 \left( 10\ \MeV/m_X \right)^{-2}$.

  \item \textbf{Constraint ``B'':}

    This constraint applies for the large width case. In contrast
    with constraint A, taking $(m_\phi,g_e) = (1\ \MeV, 1.3 \times
    10^{-9})$ is in conflict with the condition of thermal equilibrium
    between the two sectors if the DM mass $m_X \gtrsim 5\
    \MeV$. Furthermore, for $m_X \gtrsim 20\ \MeV$, the region
    $(m_\phi \sim 20 \ \MeV, g_e \sim 3\times 10^{-8})$ opens
    up. These competing effects lead to the kink in line B.

  \item \textbf{Constraint ``C'':}

    For $m_X \gtrsim 8\ \MeV$, then supernova and beam dump
    constraints allow a region of larger $g_e$: for example, $(m_\phi
    \sim 8\ \MeV, g_e \sim 6 \times 10^{-4})$ is now allowed. The red
    dashed \textit{lower bound} on $g_e$ in the left panel of
    Fig.~\ref{dd_electron_detail} then gives rise to the constraint
    ``C''. The lower bound on the cross section here comes from
    setting $m_\phi \sim m_X$, applying the red dashed lower bound on
    $g_e$, and setting $\alpha_X$ to its minimum value from CMB
    constraints.

\end{itemize}
We make two final notes.  First, in the heavy mediator case, the beam
dump constraints do not apply and the CMB constraints are in general
much stronger.  As a result, the high $\sigma_e$, low $m_X$ region
which is excluded in the light mediator case is again allowed
indicated by the light red shaded region in Fig.~(\ref{dd_SI}).
Second, if we remove the constraint $m_\phi > 1 \mbox{ MeV}$, $\phi$
will decay invisibly, and only the supernova constraints are relevant.
Then a small region of parameter space with $g_e \sim 1.3 \times
10^{-9}$ and $m_\phi < 1\ \MeV$ opens up, as discussed above under
constraint ``A.''

\begin{figure}
     \begin{center}
     \includegraphics[width=.47\textwidth]{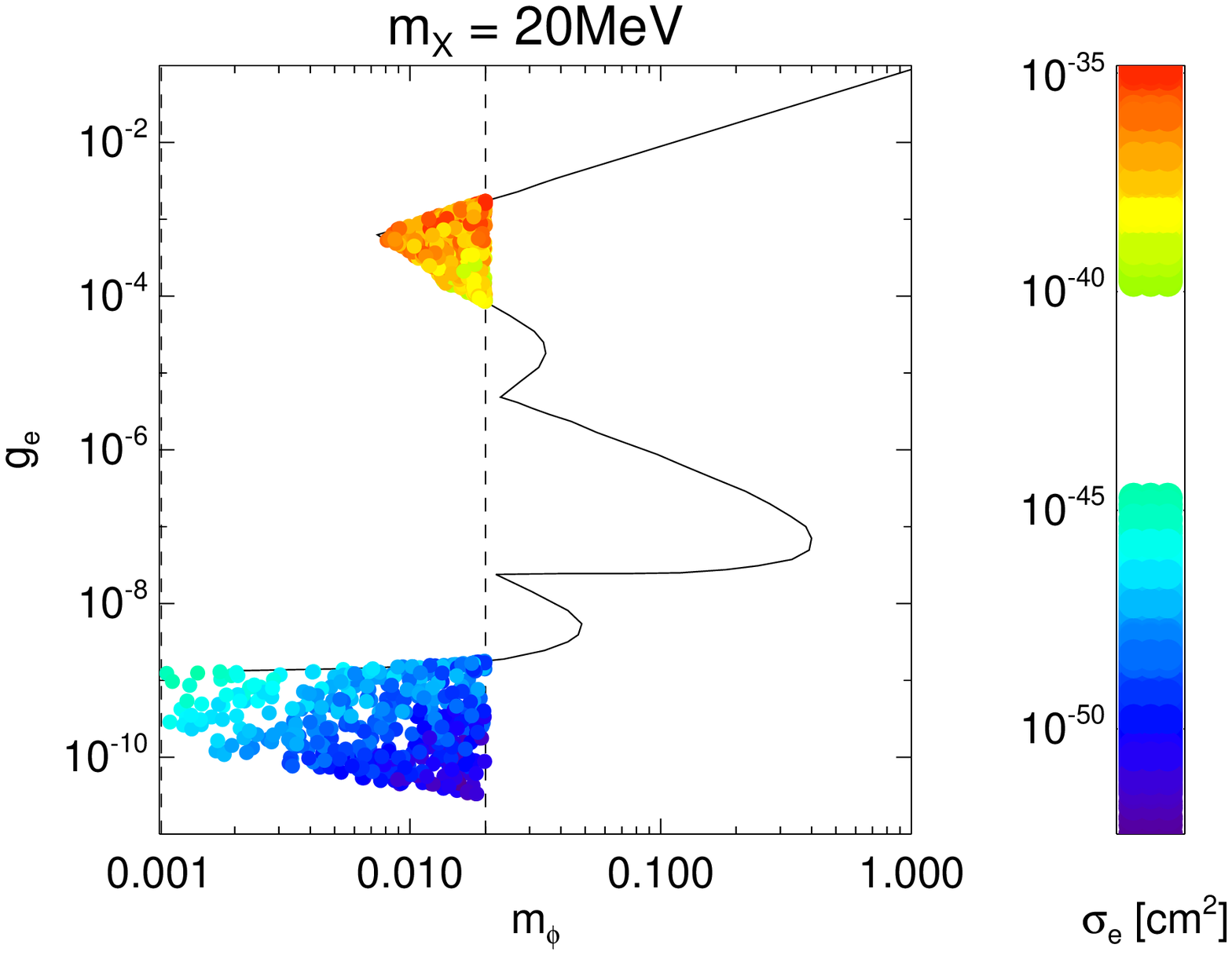}
     \includegraphics[width=.47\textwidth]{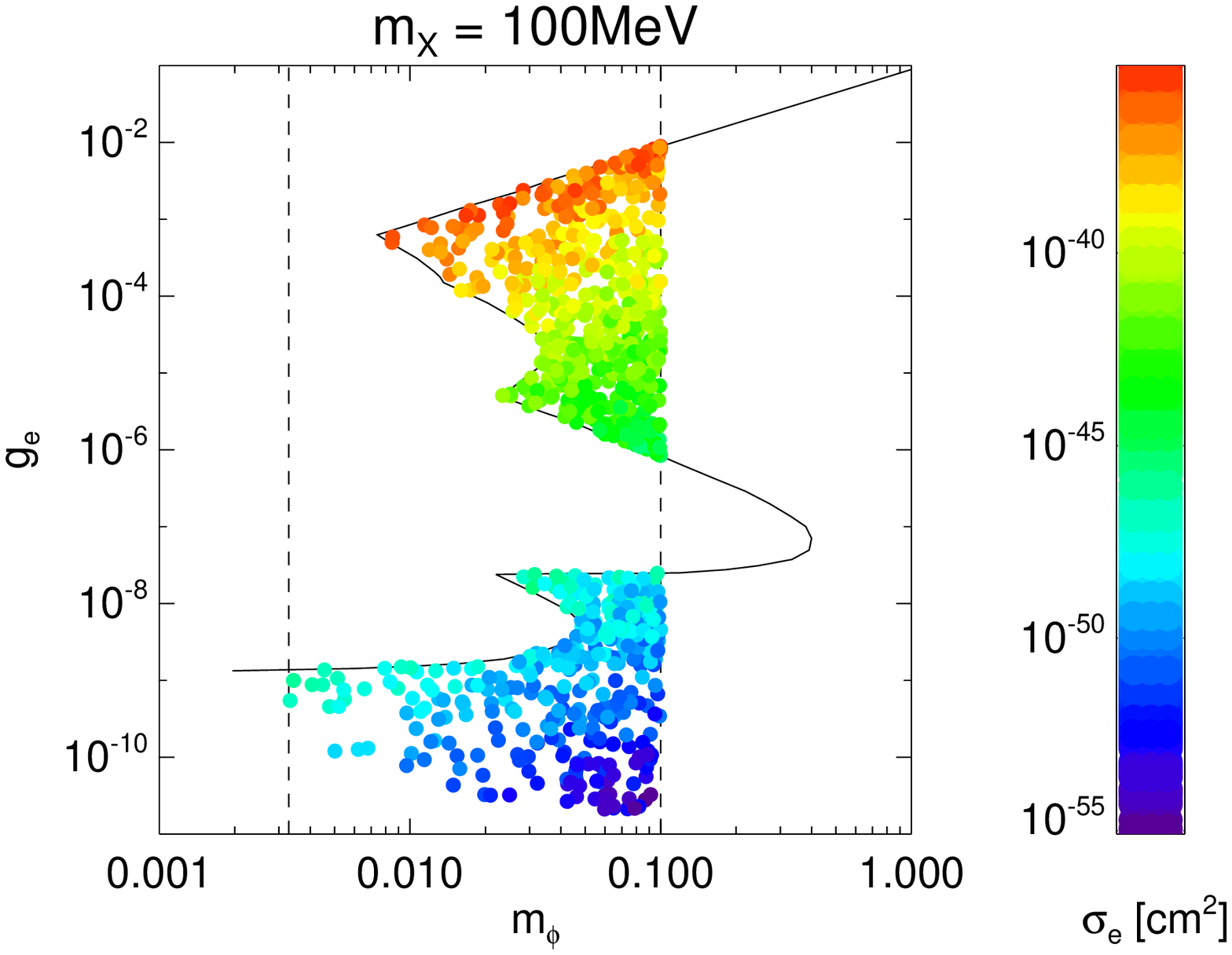}
     \caption{\label{electron_gensigma}For fixed $m_X$ and a
     mediator with mass $m_\phi < m_X$, we generate random values of $(m_\phi,
     g_e)$ allowed by beam dump, supernova, $a_e$, and BBN
     constraints. We show a sample of allowed points in the $(m_\phi,
     g_e)$ parameter space; the solid curve is extrapolated from the
     constraints in \cite{Bjorken:2009mm}, also shown in
     left panel of Fig.~(\ref{dd_electron_detail}). For each $(m_\phi,
     g_e)$ point, we then
     sample the allowed $\alpha_X$ satisfying halo shape and relic
     density constraints, and compute the corresponding elastic
     scattering cross section $\sigma_e$. The \textit{color} of the
     point is determined by $\sigma_e$. \textit{(Left)} $m_X = 20\ \MeV$,
     where the minimum mediator mass is $m_\phi = 1\
     \MeV$. \textit{(Right)} $m_X = 100\ \MeV$, where the minimum
     mediator mass $m_\phi \gtrsim 3\ \MeV$ is set by halo shape
     constraints.  }
          \end{center}
\end{figure}

We have verified the bounds discussed above by performing a general
scan of the hidden sector parameter
space. Fig.~(\ref{electron_gensigma}) illustrates our method. We begin
by mapping out the parameter space of $(\mphi,g_e)$ and require either
large $\phi$ width or $\phi$ decay before BBN. We combine this with
the constraints in \cite{Bjorken:2009mm}, given by the solid curve in
the top panels of Fig.~(\ref{electron_gensigma}). In doing so, we
impose the limit $1\ \MeV < m_\phi < m_X$ for the case of $\mphi<\mx$
and $m_\phi > 2 m_X$ in the case where $\mphi\gg\mx$. The lower limit
of $m_\phi > 1\ \MeV$ is imposed in order to allow for $\phi$ decay to
electrons. If the halo shape constraint gives a stronger lower bound
on $m_\phi$, then we take $(m_\phi)_{\textrm{min,halo}} < m_\phi <
m_X$ for the $\mphi<\mx$ case, where $(m_\phi)_{\textrm{min,halo}}$ is
minimum mediator mass allowed by the halo shape constraint. This
generates the sampled points in $(\mphi,g_e)$ that we have shown.  For
a fixed $(\mphi,g_e)$, a range of values for $\alpha_X$ is allowed,
giving rise to a range of allowed scattering cross sections.  We
sample random $\alpha_X$ values, subject to the halo shape constraint
and the relic density constraint as in \eq{eq:electron_minalphaX}.
This then gives a randomly sampled $\sigma_e$ value, which we indicate
by the color of the point in Fig.~(\ref{electron_gensigma}). For a
fixed $m_X$ value, because of the range of allowed $m_\phi$ and
$\alpha_X$ values, excluded regions in $g_e$ do \textit{not} directly
map to an excluded region in $\sigma_e$. An excluded region in
$\sigma_e$ only arises if a sufficiently large region of $g_e$ is
excluded, as shown in the left plot of
Fig.~(\ref{electron_gensigma}). We thus verify the possible values of
$\sigma_e$ in this way, imposing all the constraints
self-consistently.

\section{Conclusions}

Given the unknown nature of DM, it is important to carry out
broad-based studies of models of DM. In this paper, we have examined
constraints on thermal DM with mass $1~{\rm MeV}\lesssim \mx
\lesssim10~{\rm GeV}$, a mass range interesting for numerous
phenomenological and theoretical reasons.  We considered bounds from
cosmology, colliders and astrophysics, and derived implications of
these constraints on direct detection.

CMB constraints on DM annihilation present the most serious
challenge for light thermal DM, excluding symmetric thermal
relic DM with $s$-wave annihilation if $m_X \lesssim 1-10\
\GeV$. Two natural ways to evade this constraint are to have a DM
number asymmetry or velocity suppressed annihilation. In the
asymmetric case, we found the constraint on the annihilation cross
section such that the symmetric component efficiently annihilates
away; the minimum cross section is larger than the usual thermal relic
cross section by a factor of a few, depending on the mass. 

Achieving this minimum cross section is difficult if annihilation
occurs through a weak scale (or heavier) mediator.  Collider and
direct detection constraints have forced the presence of relatively
light mediator states in the hidden sector in order to achieve the
correct relic abundance and evade the CMB bounds. On the other hand,
we found that the DM halo shape bounds on DM self-interactions require
that the mediator is not too light. We examined constraints from
elliptical galaxy NGC720 and elliptical clusters,
and derived a lower bound on the mass of the mediator particle.

We also calculated the range of scattering cross sections allowed
within this scenario. Although the lowest bound which is
cosmologically consistent is well below the reach of any current or
envisioned direct detection experiments, we showed that several
cosmologically interesting benchmarks could be reached.  For example,
in the case of scattering off nucleons, a hidden sector in thermal
contact with the SM at $T \sim m_X$ can be ruled out if an experiment
can reach cross sections with $\sigma_n \lesssim 10^{-48} \mbox
{cm}^2$. In the case of scattering off electrons, the scenario where
$m_\phi \gg m_X$ can be probed by direct detection. Beam dump and
supernova constraints carve out a significant fraction of the
available parameter space if $m_\phi < m_X$.


\begin{acknowledgements}

We thank Doug Finkbeiner, Manoj Kaplinghat, Lisa Randall, Brian Shuve,
Tracy Slatyer, Luca Vecchi and Tomer Volansky for useful
discussions. HBY and TL thank the Theory Division of CERN, and HBY and
KMZ thank the Aspen Center for Physics where part of work was
done. The work of TL was partially supported by NASA Theory Program
grant NNX10AD85G.  The work of HBY and KMZ was supported by NASA
Theory Program grant NNX11AI17G and by NSF CAREER award PHY 1049896.

\end{acknowledgements}

\bibliography{adm}

\end{document}